\documentclass{aastex631}

\shorttitle{Radiative MHD simulation of active region scale flux emergence}
\shortauthors{Chen, Rempel, \& Fan}

\usepackage{natbib}
\usepackage{amsmath}
\usepackage{graphicx,subfigure}
\usepackage[percent]{overpic}
\usepackage{color}
\hypersetup{linkcolor=red,citecolor=blue}

\newcommand{\sectref}[1]{Section\,\ref{#1}}

\newcommand{\figref}[1]{Figure\,\ref{#1}}

\newcommand{\equref}[1]{Equation\,(\ref{#1})}

\begin{document}


\title{A Comprehensive Radiative Magnetohydrodynamics Simulation of \\ Active Region Scale Flux Emergence from the Convection Zone to the Corona}

\author[0000-0002-1963-5319]{Feng Chen}
\affiliation{School of Astronomy and Space Science, Nanjing University, Nanjing 210023, China; chenfeng@nju.edu.cn}
\affiliation{Key Laboratory for Modern Astronomy and Astrophysics (Nanjing University), Ministry of Education, Nanjing 210023, China}
\affiliation{Laboratory for Atmospheric and Space Physics, University of Colorado Boulder, Boulder, CO 80303, USA}
\affiliation{High Altitude Observatory, National Center for Atmospheric Research, Boulder, CO 80307, USA}
\author[0000-0001-5850-3119]{Matthias Rempel}
\affiliation{High Altitude Observatory, National Center for Atmospheric Research, Boulder, CO 80307, USA}
\author[0000-0003-1027-0795]{Yuhong Fan}
\affiliation{High Altitude Observatory, National Center for Atmospheric Research, Boulder, CO 80307, USA}

\correspondingauthor{Feng Chen}
\email{chenfeng@nju.edu.cn}


\begin{abstract}
We present a comprehensive radiative magnetohydrodynamic simulation of the quiet Sun and large solar active regions. The 197\,Mm wide simulation domain spans from 18 (10)\,Mm beneath the photosphere to 113\,Mm in the solar corona. Radiative transfer assuming local thermal equilibrium, optically-thin radiative losses, and anisotropic conduction transport provide the necessary realism for synthesizing observables to compare with remote sensing observations of the photosphere and corona. This model self-consistently reproduces observed features of the quiet Sun, emerging and developed active regions, and solar flares up to M class. Here, we report an overview of the first results. The surface magnetoconvection yields an upward Poynting flux that is dissipated in the corona and heats the plasma to over one million K. The quiescent corona also presents ubiquitous propagating waves, jets, and bright points with sizes down to 2\,Mm. Magnetic flux bundles emerge into the photosphere and give rise to strong and complex active regions with over $10^{23}$\,Mx magnetic flux. The coronal free magnetic energy, which is approximately 18\% of the total magnetic energy, accumulates to approximately $10^{33}$\,erg. The coronal magnetic field is clearly non-force-free, as the Lorentz force needs to balance the pressure force and viscous stress as well as drive magnetic field evolution. The emission measure from $\log_{10}T{=}4.5$ to $\log_{10}T{>}7$ provides a comprehensive view of the active region corona, such as coronal loops of various lengths and temperatures, mass circulation by evaporation and condensation, and eruptions from jets to large-scale mass ejections.
\end{abstract}


\section{INTRODUCTION}\label{sec:intro}
The magnetic field of the Sun emerges from its interior to its surface \citep{Zwaan:1985,Harvey+Zwaan:1993} and further to the Sun's atmosphere \citep{Low:1996}. The magnetic field plays an important role in the coupling between the Sun's interior and atmosphere and even beyond \citep{Solanki+al:2006,Wedemeyer+al:2009}. Active regions are formed through the emergence of large-scale strong magnetic flux and give rise to solar eruptions such as flares and coronal mass ejections \citep{vanDriel+Green:2015,Toriumi+Wang:2019}.

Magnetohydrodynamic (MHD) simulations have been the most common tool for modeling flux emergence, particularly through the upper convection zone into the solar atmosphere \citep[e.g.,][]{Fan:2001,Archontis+al:2004,Manchester+al:2004,Magara:2006,Isobe+al:2006,Arber+al:2007,Archontis+Toeroek:2008,Hood+al:2009,Fan:2009,MacTaggart+Hood:2009,MacTaggart+Haynes:2014,Leake+al:2013,Leake+al:2014,Fang+Fan:2015,Takasao+al:2015.emer,Toriumi+Takasao:2017}. More comprehensive reviews on numerical models of flux emergence and their connection with observations are given by  \citet{Hood+al:2012,Cheung+Isobe:2014}, \citet{Fan:2021.review}, and \citet{Schmieder+al:2014}.

The aforementioned examples considered the background stratification in the volume spanning from the upper convection zone to the solar atmosphere but employed idealized treatments on the thermodynamics of plasma. In addition, convective motions that pose non-negligible impacts on the evolution of the magnetic field \citep{Nordlund+al:2009,Stein:2012.review} were also missing. A fully self-consistent model of the interaction between magnetic field and plasma requires a more realistic representation of plasma thermodynamics, which is also the basis for forward synthesis of plasma emission that can be quantitatively compared to remote-sensing observations. For this purpose, it becomes necessary to account for the important physical processes that dominate the evolution of thermodynamic properties of plasma in the real Sun, such as radiative and conductive transport, energy dissipation, and ionization effects.

A fully self-consistent and realistic model for the whole domain from the convection zone to the corona is a very challenging task,  particularly if a large horizontal domain and active-region-scale magnetic flux are also desired. Tremendous efforts have been made to tackle the problem from different aspects. A  viable approach is dividing the problem into two parts: the convection zone to the photosphere and the photosphere to the corona.

Radiative MHD (RMHD) simulations that solve the radiative transport problem for optically thick radiation together with MHD equations became available for realistic simulations of flux emergence from the convection zone to the photosphere. For example, \citet{Cheung+al:2007,Cheung+al:2008} simulated the emergence of small-scale flux tubes, and \citet{Tortosa+Moreno:2009} and \citet{Moreno+al:2018} studied flux emergence on the scale of granulations. These simulations showed good agreement with observations. RMHD simulations of the uppermost convection zone were extended for problems on larger scales, such as the formation of large active regions through the emergence of coherent flux tubes \citep{Cheung+al:2010,Rempel+Cheung:2014} or by imposing a uniform flux sheet \citep{Stein+al:2011,Stein+Nordlund:2012}. These studies pointed to the important role of convection in the evolution of emerging magnetic flux. To study the relation between the solar surface and the deeper convection zone, \citet{Chen+al:2017} studied the coupling between a convective dynamo simulation down to the bottom of the convection zone and a simulation of the upper 30\,Mm. Recently, \citet{Toriumi+Hotta:2019} and \citet{Hotta+Iijima:2020} demonstrated simulations of flux emergence in an unprecedented domain that covers the full depth of the convection zone.

For the domain spanning from the photosphere to the corona, the primary problem in a realistic model is the generation and transport of energy flux into the upper atmosphere and the dissipation of energy, which can maintain the high temperature of the corona. \citet{Gudiksen+Nordlund:2002,Gudiksen+Nordlund:2005a,Gudiksen+Nordlund:2005b} presented the first realistic model of the solar corona that considers the balance between the energy losses through optically thin radiation, thermal conduction, and heating through dissipation of magnetic energy. The Poynting flux resulting from photospheric magneto-convection has been found to be sufficient to heat the corona to over one MK. Forward synthesis of extreme ultraviolet (EUV) emission \citep{Peter+al:2004,Peter+al:2006} demonstrated that coronal loops and plasma dynamics in this realistic model are consistent with observations. This method was further extended by incorporating magnetic and velocity fields from observations. \citet{Bourdin+al:2013}  found that loops in the synthetic observation appear at the same location as the observed loops and reproduced the observed Doppler shift patterns. \citet{Warnecke+Peter:2019}  showed that the model synthesized EUV emission reproduces many features observed in the very active region whose photosphere observation was used to drive the model, such as long loops connecting the spots, fan-like loops at the active region boundary, and compact loops in the active region core.

Combining the modeling methods that have been successfully applied to the upper convection zone and the atmosphere in a single computational domain is a very non-trivial step. A larger vertical extension is a clear factor to increase the computation expense. Some realistic treatments on the physical processes, such as solving radiative transfers through multiple rays, are intrinsically computationally expensive. The situation becomes worse because more stringent constraints on time-stepping are imposed by including more physics, for example, thermal conduction in the corona.

As alternative approaches, \citet{Chen+al:2014,Chen+al:2015} tried to couple a realistic coronal simulation to a realistic flux emergence simulation \citep{Rempel+Cheung:2014} and studied the evolution of active region scale magnetic flux into the corona. The behavior of coronal loops is found to be closely related to the upward energy flux generated by the forming sunspots. \citet{Abbett:2007} presented a model of the convection zone and corona in a single volume. Using artificial source terms for the optically thick radiation in the lower atmosphere and an empirical heating model for the corona, the model yields thermodynamic structures and convective patterns similar to those observed on the Sun. \citet{Fang+al:2010,Fang+al:2012} applied the same approximation to study flux emergence and investigated the impact of convective flows on emerging magnetic flux tubes.

Fully self-consistent and realistic simulations have arisen over the past decade. Many of these simulations further considered non-local thermal equilibrium (NLTE) effects that are important for modeling the chromosphere \citep{Carlsson+al:2019}. \citet{Wedemeyer+al:2012}  showed that simulated vortex motions driven by magneto-convection can reproduce and explain chromospheric swirl events and their counterparts in the transition region and lower corona. \citet{Iijima+Yokoyama:2015}, \citet{DePontieu+al:2017}, and \citet{Martinez+al:2017.science} studied small-scale activity, such as jets and spicules and their role in mass and energy transport through the atmosphere. \citet{Hansteen+al:2010,Hansteen+al:2015} presented coronal heating models, in which the generation of energy flux by surface convection and the heating of corona through dissipation is self-consistently considered in a single volume.  \citet{Martinez+al:2019} found the conversion of kinetic energy by shocks to be an important contribution to the magnetic energy in the chromosphere. The emergence of small-scale magnetic flux tubes and the response of the atmosphere were studied by \citet{Martinez+al:2008}. \citet{Archontis+Hansteen:2014} and \citet{Hansteen+al:2017} revealed that reconnection, which may occur between emerging and preexisting magnetic fields or between emerging bipoles, gives rise to Ellerman bombs, UV bursts, and small-scale flares in the lower solar atmosphere.

Due to computational expense, the aforementioned (fully self-consistent and realistic) simulations were limited to setups with a relatively small spatial domain more representative of the quiet Sun or plage regions. A major difficulty for extending these simulations to active region scales (for both the spatial domain extent and the amount of magnetic flux) is the Alfv\'en velocity that can reach unrealistically high values in the atmosphere above strong sunspots. \citet{Rempel:2017} extended the MURaM code \citep{Voegler+al:2005} with coronal physics, using a computational efficient treatment that limits the Alfv\'en velocity through the Boris correction and treats heat conduction in a hyperbolic approach (see Section \ref{sec:method} for further detail). Using this approach, \citet{Rempel:2017} simulated a stable active region comprised of a pair of strong sunspots. Synthetic EUV emission shows loop-like features arching over the corona between the two sunspots, which is in line with previous realistic simulations that consider the domain above the photosphere. This method also allows numerical simulations to tackle more extreme conditions during solar eruptions with affordable computation expense. \citet{Cheung+al:2019} presented the first realistic simulation of a full solar flare lifecycle, in which the simulated flare is triggered by the emergence of a small flux tube next to a large sunspot in a pre-existing bipolar active region.

This recent progress in realistic simulations of active regions and flares and in combination with ever increasing computational resources allows us to expand this work by developing comprehensive active-region scale flux emergence from the uppermost convection zone into the corona. The new simulation considers an unprecedented combination of a large spatial domain, a long temporal evolution, and a high degree of realism, and provides a comprehensive model of the quiet Sun, flare productive active regions, and solar eruptions (flares and CMEs) in a continuous evolution. Compared with previous simulations, the new quiet Sun model has a much larger and higher domain; the new active region model covers the whole evolution stage from emergence to giving rise to eruptions; and the new flare catalog is extended from one single event to over 100 events with largely different characteristics. The aim of this paper is to introduce the new simulation and present some highlights from a preliminary analysis of the data. Detailed analysis of the dataset for specific topics and in-depth discussions with existing theories and observations are beyond the scope of this paper and are more appropriate for more focused studies in the future. The rest of the paper is organized as follows. The numerical method and setup of the simulation are described in \sectref{sec:method}. \sectref{sec:res_qs} presents the general properties of the quiet Sun and discusses the energy flux for coronal heating. \sectref{sec:res_ar} presents the properties of the coronal magnetic field during flux emergence into the solar atmosphere. \sectref{sec:res_em} presents the dynamics of the active region corona in three evolution stages. We summarize the results in \sectref{sec:sum} and conclude in \sectref{sec:con}. 

\section{NUMERICAL SIMULATION}\label{sec:method}

\begin{figure*}
\center
\includegraphics{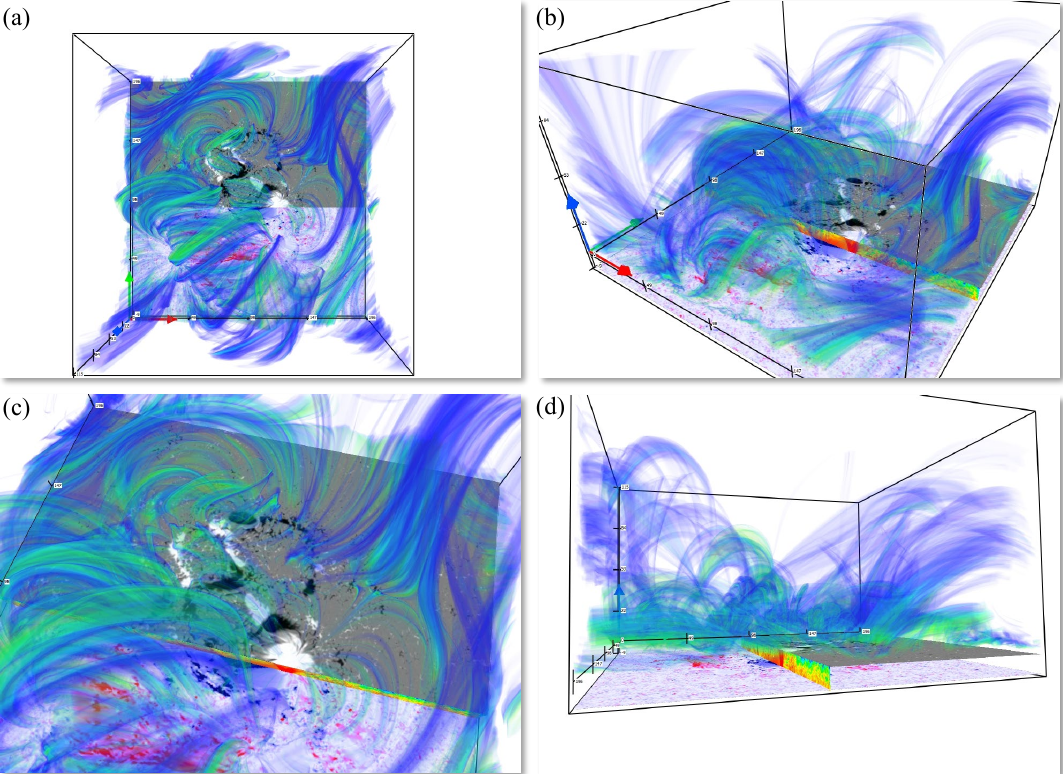}
\caption{A 3D visualization of the AR run at $t{=}$27\,h\,21\,m\,38\,s. The red, green and blue arrows indicate the $x$, $y$, and $z$ directions, respectively. The red-blue image at the bottom of the frame shows $B_{x}$ at the bottom boundary. The grayscale image covering half of the horizontal domain shows $B_{z}$ in the photosphere. The short vertical $x$--$z$ cut at the middle of the $y$ axis shows the magnetic field strength beneath the sunspot. The  green and blue colored features spreading in the volume illustrate plasma at temperatures of  2 and 3\,MK, respectively (plasma at other temperatures is made transparent in this illustration.). This figure is produced with VAPOR \citep{vapor}. An animation available in the electronic version of the paper shows an overview of the domain with the point-of-view moving from the top to the side of the domain.  
\label{fig:model_3d}} 
\end{figure*}

\subsection{Physics in the Simulation}\label{sec:phys}
In this study, we conduct numerical simulations with the {\it MURaM} radiative MHD code \citep{Voegler+al:2005,Rempel:2014}, which has been extended for modeling the solar corona \citep{Rempel:2017}. The code solves in a Cartesian domain a set of equations for the induction of magnetic field and the conservation laws of mass, momentum, and plasma energy, which is the sum of the internal and kinetic energy of plasma. The energy equation also considers sophisticated treatments on the physics that are important for thermodynamics of the plasma in the solar atmosphere. These include

\begin{itemize}
\item Radiative transfer in the optically thick regime solved along 24 rays under local thermal equilibrium approximation\citep{Voegler+al:2005},

\item Energy loss by optically thin radiation in the transition region and corona \citep[CHIANTI database][]{Landi+al:2012},

\item Anisotropic thermal conduction along magnetic field lines for fully ionized plasma \citep{Spitzer:1962},

\item  Tabular equation-of-state, which is a merger between OPAL \citep{OPAL_eos} and Uppsala \citep{Uppsala_eos} equation-of-state tables \citep[provided by the BIFROST team][]{Gudiksen+al:2011}.
\end{itemize}

The chromosphere in the simulation presented here is treated in a simplified manner by assuming local thermal equilibrium (LTE), i.e., the radiative transfer and EOS are solved by the same approach as for the photosphere. Radiation of plasma hotter than $2\times10^4$\,K is modeled by the optically thin radiative loss.  A more sophisticated treatment that considers non-LTE effects in the chromosphere, such as in the BIFROST code and a new version of the MURaM code \citep{Przybylski+al:2022}, would be computationally too demanding for simulations on the spatial and temporal scales presented here. The effects of partial/non-equilibrium ionization and ion-neutral interactions \citep{Cheung+Cameron:2012,Danilovic:2017,Khomenko+al:2018,Martinez+al:2012,Martinez+al:2020} are not accounted for in our simulation.

The other major challenges in simulations that contain a corona are the stringent constraints on the timestep from large Alfv\'en velocities, particularly in the low density areas above sunspots, and the large field-aligned thermal conductivity in the corona.

The corona extension of MURaM \citep{Rempel:2017} introduces the so-called \emph{Boris correction} \citep{Boris:1970:BC,Gombosi:etal:2002:SR} to limit the Alfv\'en velocities. This approach uses a semi-relativistic form of the MHD equations, where the speed of light imposes an upper limit on the Alfv\'en velocity.   The maximum allowed Alfv\'en velocity is dynamically adjusted to ensure that it is sufficiently larger than the sound speed ($c_s$) and velocities of plasma dynamics ($\mathbf{v}$) and is given by $\max(2 c_s, 3\vert\mathbf{v}\vert)$. The effects of this dynamical formulation have been tested in other MURaM simulations. \citet{Rempel:2017} compared the solutions with different (fixed) Alfv\'en velocity limits for a non-flaring active region setup, where the horizontally averaged Alfv\'en velocity is on the order of $10^{4}$\,km/s, and found quantitatively similar model coronae. In a simulation of a C-class flare, \citet{Cheung+al:2019} found comparable results in the reference run, which used a dynamically adjusted Alfv\'en velocity limit as in the present simulation and a test run, which used a three times higher limit (${\approx}3\times10^4$\,km/s). Furthermore, the Boris correction does not change the force balance in the momentum equation, only the time-scales at which the system reacts to an imbalance. This means that the effective $\beta$ value of the corona is not changed. The corresponding term of the Boris correction in the energy equation \citep[Equation (30) in][]{Rempel:2017} is monitored in our simulation and found to be small. Therefore, we consider the impact of the Boris correction to be negligible. If waves (in strong active regions) are a primary research interest, a rerun of a short period of the simulation could be done with a fixed high Alfv\'en velocity limit.

An approach with a similar concept is also applied to anisotropic thermal conduction. In addition to the MHD equations, the code solves a hyperbolic equation (i.e., non-Fickian) for the field-aligned thermal conduction flux \citep{Gombosi:1993:telegraph,Snodin:etal:2006:non-fickian}. This approach filters out unphysical transport velocities that are present in the parabolic system and can limit the timestep substantially. Based on the expression for flux-limiters such as \citet{Fisher:1986}, it can be estimated that the free-streaming limited transport velocity reaches about twice the ion speed of sound. The maximum Alfv\'en velocity ($\max(2 c_s, 3\vert\mathbf{v}\vert)$) ensures a proper treatment of free-streaming limited heat conduction. The implementation and choice of parameters for the Boris correction and hyperbolic thermal conduction are discussed in \citet{Rempel:2017}.

In summary, the extended version of the MURaM code considers a compromise between physical complexity and numerically efficient treatment that enables comprehensive simulations in a single domain that spans from the upper-most convection zone into the corona.

\subsection{Simulation Setups}\label{sec:model}
We conduct the simulation with two stages: quiet Sun (QS) and active region (AR). The quiet Sun stage aims to establish a solar atmosphere that reaches a dynamic equilibrium. It is not only a stand-alone model for studies of coronal heating, energy and mass circulation, and small-scale dynamics in the quiet Sun but also the initial condition for the active region stage. The active region stage considers a large amount of magnetic flux emerging on a spatial scale of a hundred Mm and focuses on the evolution of the magnetic field, emission structures in the active region corona, and solar eruptions.

For the description of the model setup and results, we first define the notations used throughout the rest of the paper. $x$ and $y$ are the two horizontal directions of the domain, and $z$ is the vertical direction. $L_{x}$, $L_{y}$, and $L_{z}$ are the lengths of the domain in each direction. $N_{x}$, $N_{y}$, and $N_{z}$ are the number of grid points in each direction. $\Delta x$, $\Delta y$, $\Delta z$ are the grid spacings. In particular, $z={0}$ is defined by the position of the photosphere, the layer in which the average (Rosseland) optical depth ($\tau$) equals unity.

\subsubsection{Quiet Sun Stage}\label{sec:model_qs}
The domain for the quiet Sun stage (hereafter, QS run) covers an area of $L_{x}\times L_{y}{=}196.608^{2}$\,Mm$^2$. The total vertical extent ($L_{z}$) is 131.072\,Mm, with about 18\,Mm beneath the photosphere. The domain is resolved by $N_{x}\times N_{y}\times N_{z}{=}1024\times1024\times2048$ grid points with a grid spacing of $\Delta x \times \Delta y \times \Delta z{=}192^{2}\times64$\,km$^{3}$.

The chosen grid spacing helps to achieve better computation efficiency (i.e., larger timesteps) while still capturing the scale of granulation and the atmospheric scale-height near the photosphere. The results with different grid spacings have been extensively tested in previous numerical experiments with smaller domains. A grid spacing of $\Delta x \times \Delta y \times \Delta z{=}192^{2}\times64$\,km$^{3}$ does still provide in MURaM simulations the correct photospheric energy flux, scale of granulation and convective root-mean-square velocity. Lower resolution would only capture meso-granular scales and associated energy transport.

The lateral boundaries of the domain are periodic for all variables. The bottom boundary is symmetric for mass flows and all components of the magnetic field. The top of the domain is set at about 113\,Mm above the photosphere. The magnetic field in the ghost cells  of the top boundary is a potential field calculated based on $B_z$ at the top of the domain.  The top boundary is symmetric for the horizontal velocities. We use a damped boundary for the  vertical velocity, such that the  vertical velocities in the first and second ghost cells are reduced to 50\% and 25\% of values in the last two cells of the computation domain, respectively.  The top boundary, albeit considered open, strongly suppresses mass and energy flux at and near the top of the computation domain. For example, the net transport velocity 
$$
\frac{\int \rho v_{z}dxdy}{\int \rho dx dy}
$$
in the horizontal layer at 3\,Mm below the top boundary fluctuates with an amplitude of about 0.1\,km/s. We also note that a persistent outflow of about 5\,km/s is generated in a coronal hole run (not presented in this paper), where outflows are not damped by the top boundary,.

The QS run is constructed by adding an isothermal corona to a surface magneto-convection simulation with a domain that extends from the convection zone to the upper photosphere as follows.  An 18\,Mm-deep hydrodynamic convection simulation is initialized with a stratification based on the standard solar model and a small velocity perturbation and is evolved for a sufficiently long time until convection is fully developed. After adding a small seed magnetic field, the magneto-convection simulation is further evolved until the magnetic field saturates. The magnetic field is a small-scale mixed polarity field generated by a small-scale dynamo operating in the turbulent convection zone \citep{Voegler+Schuessler:2007,Rempel:2014}. At the resolution used in this simulation, the small-scale dynamo is no longer captured in the photosphere, but the deeper layers of the box continue to maintain a small-scale field. This field is transported into the photosphere by overturning convection and does maintain a quiet Sun mixed-polarity network that is critical for the heating of the corona. The 18\,Mm depth of the convective layer allows the development of convection cells comparable to supergranulation in scale \citep{Lord+al:2014} and flux imbalance on that spatial scale (the total flux of the entire domain is balanced). To limit transients, the initialization process of the corona is done in two steps. In the first step, a layer of 10\,Mm is added and the simulation is evolved for a few hours until the transition region develops. Then the rest of the corona (about 103\,Mm) is added using an isothermal stratification with the mean transition region temperature and a potential magnetic field extrapolated from the top of the transition region.

The simulation is evolved from the initial condition to a state in which the mean temperature and density in the corona reach an equilibrium. Then, we define this time as the beginning of the production run ($t_{QS}{=}0$), and evolve the QS run further for about 12 solar hours.

\subsubsection{Active Region Stage}\label{sec:model_ar}
The active region (hereafter, AR run) has a domain that covers an area of $196.608^{2}$\,Mm$^{2}$ (identical to that of the QS run). The vertical extent is 122.88\,Mm, reaching 9.6\,Mm beneath the photosphere. The domain is resolved by $N_{x}\times N_{y}\times N_{z}{=}1024\times1024\times1920$ grid points, which leads to the same grid spacing as for the QS run.

The bottom boundary is a time-dependent boundary coupled to a solar convective dynamo simulation by \citet{Fan+Fang:2014}. The implementation of the coupling is described in detail in \citet[][hereafter, Paper\,I]{Chen+al:2017}. In brief, we extract a time series of magnetic and velocity fields in a horizontal layer (a constant radius surface in spherical coordinates) in the dynamo simulation. This time series captures the evolution of several superequipartition magnetic flux bundles that are self-consistently generated in the convective dynamo. The extracted data are used to fill the ghost cell values in the AR run in this study. The benefits of coupling are twofold. This solves the problem that global-scale convective dynamo simulations under the anelastic assumption cannot reach the solar surface. It provides a more realistic and physically consistent lower boundary flux transport from the deep interior for the near-surface flux emergence simulation. The boundary conditions for the lateral and top boundaries for the AR run are identical to those of the QS run.  The top boundary is similar to the QS run. We impose strong damping on the vertical velocity, which leads to a mostly closed boundary except for times when dynamic events such as strong flares and associated mass ejections reach the top boundary. During these times, plasma can leave the domain, while also causing partial reflection of waves \citep{WangCan+al:2021}.

We use a snapshot of the QS run at $t_{\rm QS}={5.7}$\,h as the initial condition for the AR run. When converting the QS run snapshot to the initial condition, we cut off the lowest 8\,Mm (128 grid points with $\Delta z{=}64$\,km), such that the bottom boundary of the AR run is at 9.6\,Mm beneath the photosphere and the vertical domain is resolved by 1920 grid points (compared with 2048 grid points for the QS run). In Paper\,I, we studied the effect of domain depth on the properties of the emerged flux. Using a shallower domain provides the following two advantages: (1) The time required for flux emergence through the convection zone is significantly shorter, which provides a significant saving in computing in a setup with an expensive corona. (2) A shallower convection zone  allows more magnetic flux to reach the photosphere and leads to a more complex active region that offers the prospect of a more dynamic corona.

The AR run is evolved for about 48 solar hours, which is actually limited by our computational resources. The analysis of a simulation with a similar depth for the convection zone (presented in Paper\,I) showed that the peak of the unsigned magnetic flux is reached at about 50 hours after the start of flux emergence. Therefore, this time period is sufficient to cover the emergence of magnetic flux from the convection zone to the photosphere and the formation and growth of active regions in the photosphere. This is the most dynamic stage during the evolution of active regions.

\subsection{Observable synthesis}
Differential emission measures (DEMs) describe the distribution of plasma density (along the line-of-sight) with temperature, as given by the relation 
$$
DEM(T) = n_{e}^2(T) \frac{dl}{dT},
$$
where $l$ is the length of the line of sight.
The emission measure is known as 
$$
EM(T) = DEM(T)dT.
$$
DEMs inverted from multichannel EUV/X-ray images \citep[e.g.,][]{Kashya+Drake:1998,Hannah+Kontar:2012,Cheung+al:2015} are very useful for investigating the thermal properties of coronal plasma.

To evaluate the EM in the simulation, the logarithmic temperature range starting from $\log T_{min}{=4.5}$ to the highest temperature ($\log T_{\rm max}$) in the domain (time-dependent) is discretized into $N_{em}{=}(\log T_{\rm max}-\log T_{min})/\Delta \log T$ temperature bins with an interval of $\Delta \log T{=}0.1$, i.e.,
$$
\log T_{n}{=}\log T_{min}+n\Delta \log T~~~n=0,1,...,N_{em}-1.
$$
Then, we calculate along a given line of sight the sum of $n_{e}^2$ in corresponding temperature bins by 
$$
EM_{x_{i}}(x_{j},x_{k},\log T_{n}) = \sum \limits_{m=1}^{N_{x_{i}}} n_{e}^2(m,x_{j},x_{k})f_{T}(m,x_{j},x_{k}) \Delta x_{i}, 
$$
where $x_{i,j,k}$ denote the axes of the domain \footnote{$(i,j,k){=}$(1,2,3), (2,1,3), and (3,1,2) stands for line-of-sight integrals through $x$, $y$, and $z$ directions, which yield views of the $y-z$, $x-z$, and $x-y$ planes, respectively.}. In the following we omit $x_{j}$ and $x_{k}$, which are obvious. The contribution factor $0\leq f_{T}\leq 1$ is computed based on the overlap interval method described in \citet[][Section 2.3]{Rempel:2017}.

Similarly, we also evaluate the temperature filling factor
$$
F_{x_{i}}(\log T_{n}) = \sum \limits_{m=1}^{N_{x_{i}}} f_{T}(m)\Delta x_{i}, 
$$
EM weighted line-of-sight velocity
$$
V{\rm los}_{x_{i}}(\log T_{n}) =\frac{1}{EM_{x_{i}}}\sum \limits_{m=1}^{N_{x_{i}}} v_{x_{i}}(m) n_{e}^2(m)f_{T}(m) \Delta x_{i}, 
$$
and EM weighted velocity dispersion
$$
V{\rm rms}_{x_{i}}(\log T_{n}) =\sqrt{\frac{1}{EM_{x_{i}}}\sum \limits_{m=1}^{N_{x_{i}}} v^2_{x_{i}}(m) n_{e}^2(m)f_{T}(m) \Delta x_{i}-V{\rm los}_{x_{i}}^2}.
$$
EM and EM weighted quantities are computed during runtime and available as a direct output every 50 iterations, which corresponds to a cadence of about 5 seconds in the quiet Sun, 3 seconds in typical active regions, and 0.5 seconds during strong flares in the simulation.

The prestored high cadence EM also facilitates synthesizing observations of the {\it Atmospheric Imaging Assembly} \citep[AIA,][]{AIA} on the {\it Solar Dynamics Observatory} (SDO). The intensity of optically thin radiation in the corona can be evaluated by
$$
I_{\lambda} = \int G_{\lambda}(T,n_{e})DEM(T)dT,
$$
 where $G_{\lambda}$ is a contribution function. The intensity of a particular AIA channel $I_{\lambda_{i}}$ is computed by
$$
I_{\lambda_{i}} = \int K_{\lambda_{i}}(T)DEM(T)dT,
$$
where the temperature-response function $K_{\lambda_{i}}$ ($\lambda_{i}$ denotes a filter name) is a convolution of the contribution function $G_{\lambda}$ and the wavelength-response of the instrument \citep{AIAcali}. The density dependence of $G_{\lambda}$ is weak (\citealt{Landi+Landini:1999}, see also discussions in \citealt{Peter+al:2006,Leenaarts:2020}). Although this weak dependence may still be relevant in the synthesis of spectral lines \citep[e.g., ][]{Peter+al:2006,Olluri+al:2015}, it is usually omitted in the calculation of the temperature-response function and synthetic AIA images.

We generate $K_{\lambda_{i}}$ with the {\it aia\_get\_response} procedure in the {\it SolarSoftWare} package according to the abundance applied in our simulation. AIA intensities viewed along the $x_{i}$ direction are obtained by
$$
I_{\lambda_{i}, x_{i}} = \sum \limits_{n=0}^{N_{em}-1}K_{\lambda_{i}}(\log T_{n})EM_{x_{i}}(\log T_{n}).
$$

\section{RESULTS I: THE QUIET SUN CORONA}\label{sec:res_qs}

\begin{figure*}
\includegraphics{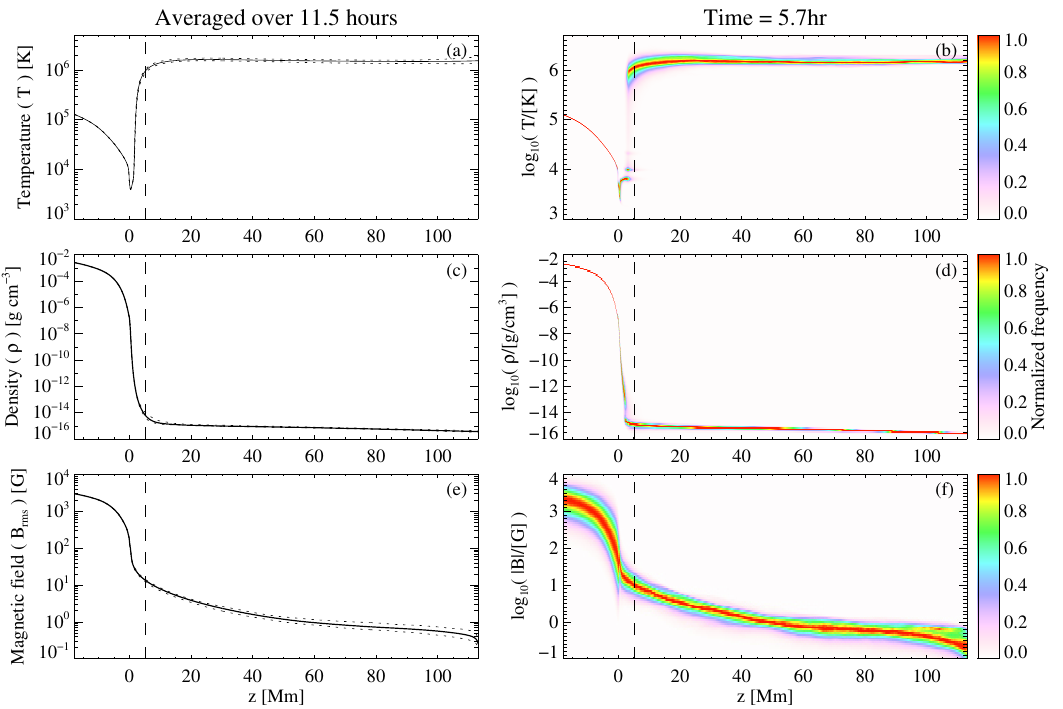}
\caption{Left column: Horizontally and temporally averaged temperature (a), density (c), and magnetic field strength (e) in the QS run. Dotted lines around the solid curves indicate the maximum and minimum values of the horizontally averaged temperature and density over the temporal evolution of about 11.5 hours. The dashed line at $z{=}5.28$\,Mm indicates the base of the corona. Right column: Probability density functions (PDFs) of the temperature, density and magnetic field strength in a snapshot at $t_{\rm QS}{=}5.7$\,h. The PDFs are evaluated and normalized to unity at each height.
\label{fig:qs_rhot}} 
\end{figure*}

\begin{figure*}
\includegraphics{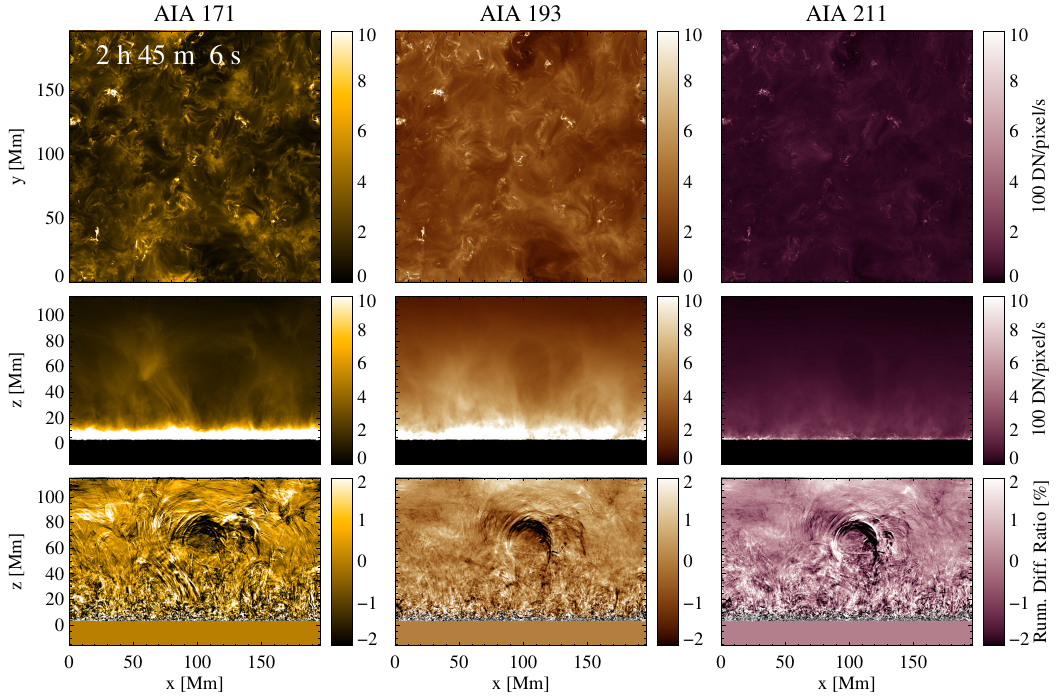}
\caption{Synthetic AIA 171, 193 and 211 channel images. The first row shows the view from the top of the domain. The middle row is a side view along the $y$-axis. The bottom row shows the running difference ratio of consecutive images in the middle row. An animation of this figure is available in the electronic version of this paper. The animation covers a temporal evolution of 50 minutes in the simulation and illustrates features such as coronal bright points, jets, campfire-like transient brightenings (\sectref{sec:res_qs_activity}), an eruption of magnetic flux rope in the quiet Sun (\sectref{sec:res_qs_fluxrope}), and ubiquitous upward propagating sound waves and Alfv\'enic disturbances (\sectref{sec:res_qs_wave}).
\label{fig:qs_multi}} 
\end{figure*}

\begin{figure*}
\includegraphics[width=18cm]{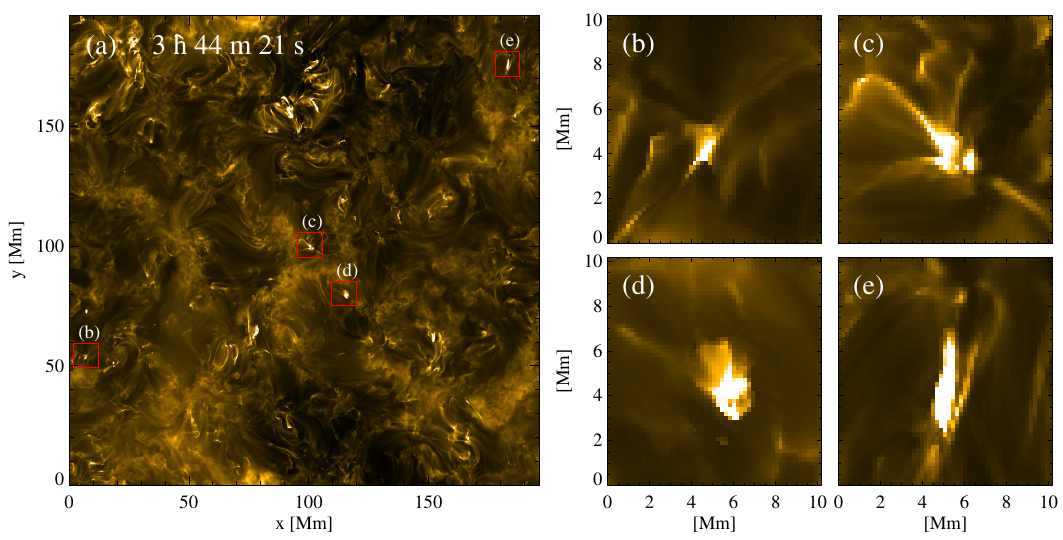}
\caption{Examples of campfire-like small-scale transient brightenings. (a): Synthetic AIA 171 image from the top view of the whole horizontal domain. Red boxes (10\,Mm wide each) mark four examples shown in a detailed view in Panels (b) -- (e), respectively. These images are displayed with the same color scale (0 -- 1000 DN/pixel/s) as in \figref{fig:qs_multi}.
\label{fig:qs_campfire}}
\end{figure*}

\begin{figure*}
\center
\includegraphics[width=18cm]{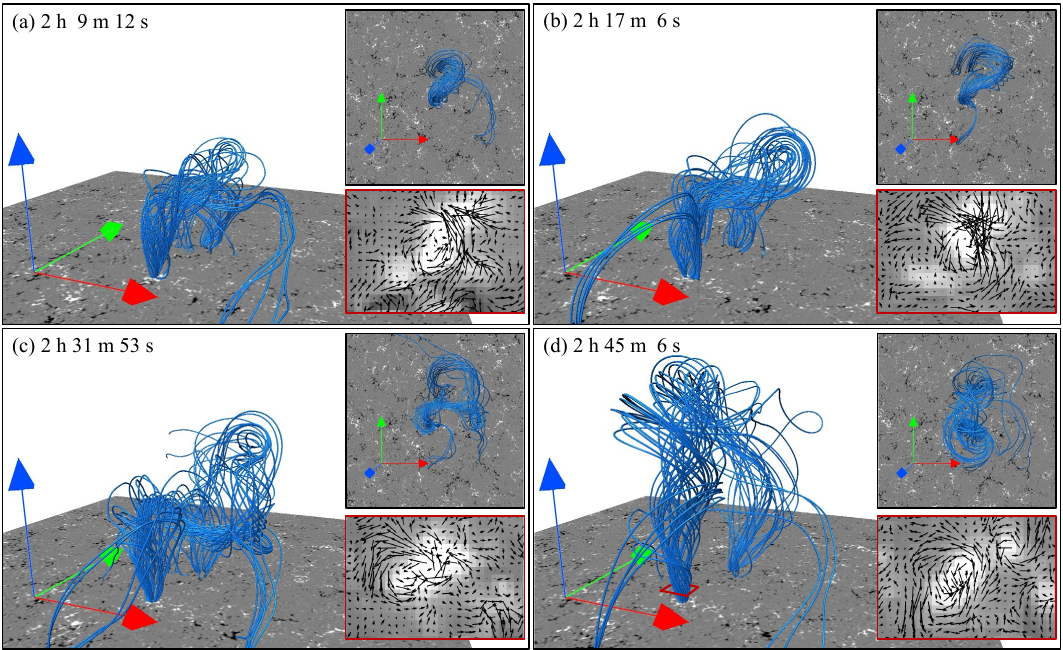}
\caption{Evolution of a magnetic flux rope spontaneously formed in the quiet Sun. The blue lines are magnetic field lines that outline the flux rope. The grayscale images in the 3D views show $B_{z}$ in the photosphere ($+/-100$\,G). Red-frame panels present $B_{z}$ (grayscale image, $+/-100$\,G) and the horizontal velocity field (arrows) at the (positive polarity) leg of the flux rope. The location of the red-frame panels (about 1\,Mm above the photosphere and fixed in time) is marked by the small red frame in Panel (d). The red, green and blue arrows indicate the $x$, $y$, and $z$ directions, respectively. The 3D visualization is produced with VAPOR \citep{vapor}.
\label{fig:qs_fluxrope}} 
\end{figure*}

\begin{figure*}
\includegraphics{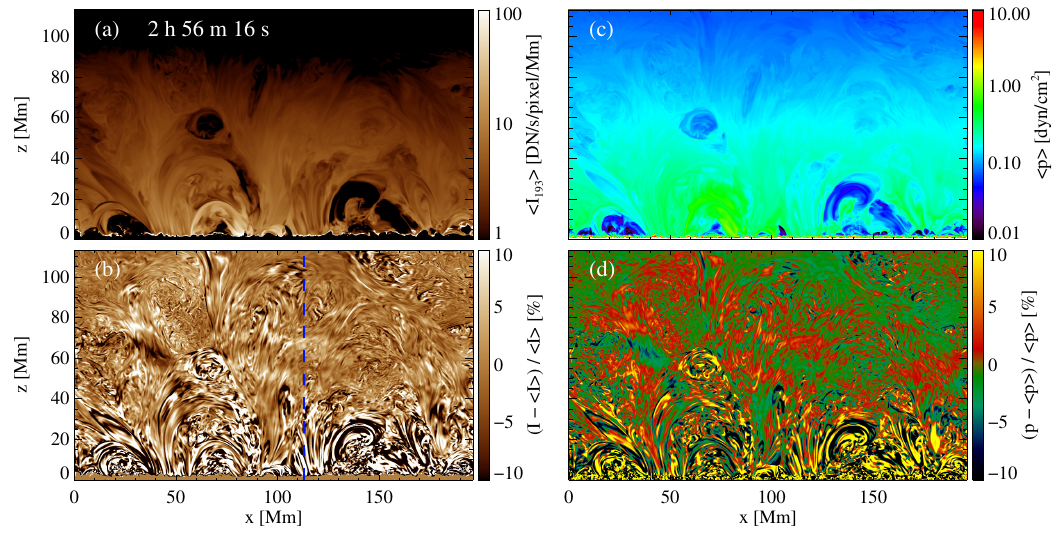}
\caption{Propagating disturbances in the quiet Sun corona. Displayed quantities are in a $x-z$ cut through the center of domain ($y=L_{y}/2$). (a): Temporally smoothed AIA 193 intensity with a 60 seconds window as defined by \equref{equ:qs_mean}. (b): Disturbances in the AIA 193 intensity on time scales shorter than the 60 seconds averaging window, as defined by \equref{equ:qs_dist}. The disturbance is displayed relative to the mean. The blue dashed line indicates the slit for the time-distance diagrams in \figref{fig:qs_timedist}. (c) \& (d): Temporally smoothed and disturbance components of plasma pressure, respectively. An animation available in the electronic version of the paper shows the evolution of the background and the upward propagating disturbances in a time period of about 70 minutes. The same time series is used to construct the time-distance diagrams in \figref{fig:qs_timedist}.
\label{fig:qs_dist}}
\end{figure*}

\begin{figure*}
\includegraphics{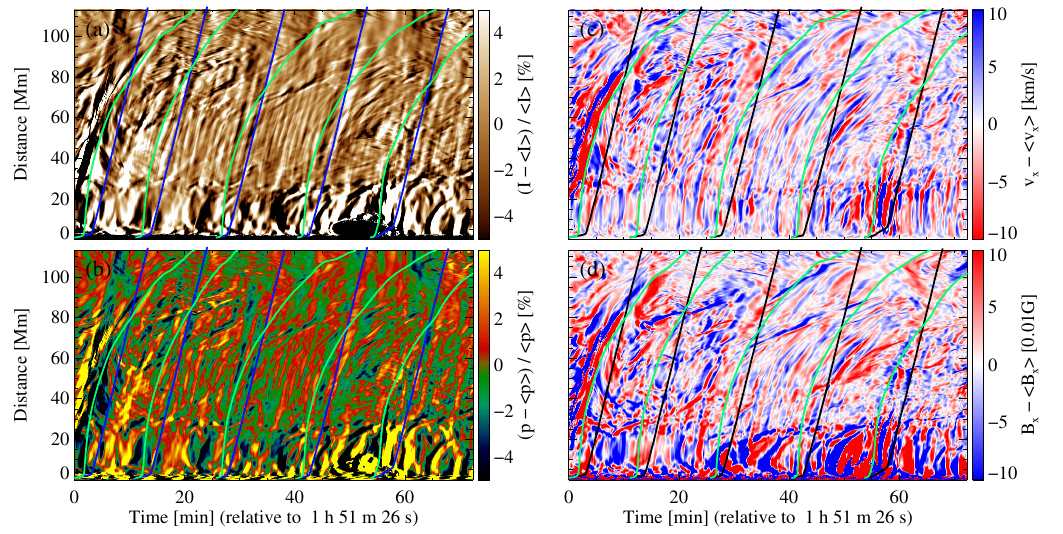}
\caption{Time-distance diagrams of the disturbances in (a) AIA 193 intensity, (b) plasma pressure, (c) $v_{x}$, and (d) $B_{x}$. The diagrams are constructed along the slit shown in \figref{fig:qs_dist}(b) and for a time period of 70 minutes. The blue lines in each panel illustrate the travel paths of sound waves (starting from 5 different time instances) according to \equref{equ:qs_pcs}. Similarly, green lines show the paths of Alfv\'enic disturbances as given by \equref{equ:qs_pa}.
\label{fig:qs_timedist}}
\end{figure*}

\subsection{Stable and Self-Maintained Quiet Sun Corona}\label{sec:res_qs_intro}
After the initial relaxation stage that has been excluded from the production run, the QS run generates a hot and tenuous corona that remains in a fully self-maintained, stable dynamic equilibrium.

We calculate the horizontally averaged temperature, density and magnetic field strength through the whole vertical domain and the temporal averages of these vertical profiles. The left column in \figref{fig:qs_rhot} shows the temporally and horizontally averaged profiles. The vertical profiles of the temperature and density reproduce a stratification consistent with the real Sun from the uppermost convective layer to the lowermost 110\,Mm corona. The magnetic field strength is about 150\,G in the photosphere and drops to below 10\,G in the corona.

Furthermore, we plot in the right column of \figref{fig:qs_rhot} the probability density functions (PDFs) of the (logarithmic) temperature ($T$), density ($\rho$), and magnetic field strength ($\vert\mathbf{B}\vert$) through the vertical domain at a given instant. The temperature and density distributions in both the interior and the high corona are very narrow, and the data points are highly aligned to the temporally and horizontally averaged profiles shown in the left column. In the transition region between $z{\approx}2$ and 4\,Mm, the temperature at the same height shows a clear bifurcation into a branch at $\log T{\approx}$4 and a branch at $\log T{\approx}$6. Meanwhile, only a small portion of this region has so-called transition region temperatures as a result of steep temperature gradients. Similar behavior can also be found in the density distribution.

The temperature and density distributions in \figref{fig:qs_rhot} also imply that the actual transition region is a very corrugated surface. To determine whether a grid point belongs to the corona, we employ the following factor ($f_{c}$):
\begin{equation}\label{equ:fcor}
f_{c}(x,y,z) = \begin{cases}
		1,~~T > 2\times10^5\,K \\
		0,~~T < 2\times10^5\,K.
	\end{cases}
\end{equation}
Hereafter, for all analyses of the QS run, we define the coronal base as the layer at $z{=}5.28$\,Mm, where the horizontally averaged $f_{c}$ first reaches 0.99. This height is marked in all panels in \figref{fig:qs_rhot} by a vertical dashed line.

We synthesize AIA images in 171, 193, and 211 channels to illustrate features and activity in the simulated quiet Sun corona. \figref{fig:qs_multi} shows synthetic images from the top and side views. In a direct comparison of the synthetic image with an actual AIA image, we find that the features seen in the synthetic image are clearly sharper. This is because the numerical simulation has a substantially higher spatial resolution (in terms of cell size compared to AIA pixel size), and we do not account for the point spread function of the instrument. 

The simulated quiet Sun corona is much more homogeneous than the corona above active regions (as shown later in this paper) and comprises large-scale structures that evolve slowly, short loops at 20--30\,Mm length, and numerous small-scale transient brightenings. 

To summarize, the formation of a corona with a realistic temperature and density indicates that the model provides sufficient energy to heat the coronal plasma to million K. Moreover, the hot corona can be maintained for over 10 hours (and more if the run continues) and is stable against activity on various time and spatial scales. This suggests a very robust balance between the heating and energy losses in the corona, as discussed later in this paper.

\subsection{Bright Points, Jets, and Campfires}\label{sec:res_qs_activity}
Small-scale explosive events and high-energy jets have been observed in the quiet Sun for decades \citep[e.g.,][]{Habbal+Withbroe:1981,Brueckner+Bartoe:1983,Dere+al:1991,Aschwanden+al:2000}. The simulated quiet Sun corona also hosts ubiquitous small-scale activity. By examining the temporal evolution shown by the animation of \figref{fig:qs_multi}, we find numerous small-scale brightening of any shape that appear and disappear on short time scales of a few minutes or tens of seconds. These features resemble coronal bright points that are commonly seen in EUV observations of the corona \citep[see, e.g.,][ and references therein]{McIntosh+Gurman:2005,Alipour+Safari:2015,Madjarska:2019}. The bright points seen in this simulation are closely related to the small-scale mixed polarity field in the quiet Sun, which is in line with the findings in observations \citep[e.g.,][]{Perez+al:2008,MouChaozhou+al:2016}. This also implies that these bright points are likely a manifestation of magnetic energy release, as suggested from both observational and theoretical bases \citep[e.g,][]{Habbal:1992,TianHui+al:2008,ZhangQingmin+al:2012,Parnell+Galsgaard:2004,Galsgaard+al:2019}.

Some of the bright points are accompanied by upsurges of plasma \citep[i.e., jets][]{Raouafi+al:2016,ShenYuandeng:2021,DePontieu:2021.IRIS}, which are illustrated more clearly by the synthetic AIA images from the side view in \figref{fig:qs_multi}. This is similar to the relation between coronal jets and bright points found in observations \citep[see, e.g.,][ and references therein]{Doschek+al:2010,ZhangQingmin+al:2016,MouChaozhou+al:2018,Panesar+al:2020}. Most of the jets in the simulation are confined below 10\,Mm in height, while some may reach a much greater height. For example the jet that occurs at about $t_{\rm QS}{=}$2\,h\,32\,m\,15\,s at $(x,y){\approx}(140,45)$ (best seen in the 171 channel in the animation) reaches approximately 40\,Mm in height with an average upward velocity of about 300\,km/s.

Recently, "campfires" as tiny brightening much smaller than typical bright points have been reported by \citet{Berghmans+al:2021}, and they are suggested to be episodes of coronal heating events. \citet{Chitta+al:2021} found that similar features, albeit in slightly larger sizes, are also common in AIA observations. \figref{fig:qs_campfire}(a) shows at an arbitrarily chosen instance a top-view synthetic AIA 171 image, in which many bright dots on spatial scales of a few Mm can be easily identified, and Panels (b)--(e) highlight a few examples among these features. They display different shapes, such as a small dot (less than 2\,Mm wide), a larger dot (over 3\,Mm wide), irregular dot with elongated branches, and an elongated dot (less than 2\,Mm wide but over 4\,Mm long), which are similar to the (larger events of) observed campfires. 

This simulation was not designed with the particular purpose of studying small events on scales of a few Mm. Nevertheless, given the five-point-stencil of the numerical algorithm and 192\,km cell size of our simulation, the campfire events are resolved by a sufficient number of grid points. In an MURaM simulation of a quiet Sun setup with a significantly higher resolution (48.8\,km cell size), \citet{ChenYajie+al:2021} found features similar to those in our simulation, which strongly supports the validity and robustness of the events we see.

\subsection{Magnetic Flux Rope eruption}\label{sec:res_qs_fluxrope}
A quiet Sun flux rope eruption is made visible by the running difference ratio of consecutive images ($(A_{i+1}-A_{i})/A_{i}$, where $A$ denotes a time series of images) in the bottom row of \figref{fig:qs_multi}. The animation associated with \figref{fig:qs_multi} covers the whole evolution of this event. The selected snapshot shows the ejection of a balloon-like structure in the running difference ratio images. It also shows a tendency of rotation and unwinding, which is clearly noticeable in the animation. Meanwhile, a dimming region is visible in the 193 channel images from both the top and side views. 

We examine the magnetic field in this region and confirm that the balloon-like structure is a magnetic flux rope that is generated spontaneously by small-scale flux concentrations and vortex motions in the photosphere. The evolution of the magnetic flux rope from build up to eruption is presented in \figref{fig:qs_fluxrope}. The structure exists long before its eruption seen in the animation of \figref{fig:qs_multi} and appears to be an already twisted flux bundle connecting two magnetic flux concentrations with a separation of about 20\,Mm, which is much larger than the size of granules and closer to that of supergranules. A persistent counterclockwise rotation (when seen from top, i.e., along $-z$) at the positive polarity (also the more coherent) flux concentration constantly adds twist to the flux bundle. At $t_{\rm QS}{\approx}$2\,h30\,m, the flux bundle becomes unstable and forms a kink shape with a dip in the middle, as shown in \figref{fig:qs_fluxrope}. The vortex motion of the positive polarity flux concentration continuously stretches the dip upward, and when seen from the side view in \figref{fig:qs_multi} (i.e., along the $y$-axis), the flux bundle rolls clockwise, as the animation of \figref{fig:qs_multi} illustrates.

Vortex motions are a common pattern in the flow field in the solar photosphere, as revealed in observations \citep[see, e.g.,][ and references therein]{Brandt:1988,Attie+al:2009,Bonet+al:2008,Bonet+al:2010,Requerey+al:2018} and in magneto-convection simulations \citep{Moll+al:2011,Shelyag+al:2011,LiuJiaJia+al:2019,Canivete+Steiner:2020}. Studies have also shown that vortex flow fields may be coupled with magnetic fields and give rise to rotating magnetic structures that extend to the atmosphere on various spatial scales \citep{ZhangJun+LiuYang:2011,Wedemeyer+al:2012,SuYang+al:2012,SuYang+al:2014,Wedemeyer+al:2013,Kato+Wedemeyer:2017,Giagkiozis+al:2018}. The flux rope formed in this simulation is a particularly intriguing case because it is a coherent magnetic structure with both legs anchoring in the photosphere. Although each of its legs is similar to a magnetic swirl/tornado, as shown in the aforementioned studies, the shape and behavior of the whole structure, albeit with a much smaller size and less magnetic flux, is more similar to flux ropes in major solar eruptions. It leaves interesting questions of how frequently these structures can form and if they can be detected.

\subsection{Propagating Disturbances}\label{sec:res_qs_wave}
The running difference ratio images in \figref{fig:qs_multi} also reveal ubiquitous propagating disturbances. They are most likely waves that are excited in the lower atmosphere and propagate upward. We briefly examine the properties of the waves in a vertical cut through the center of the domain (the $x-z$ plane at $y{=}L_{y}/2$). To better extract the wave-related disturbances from the background, we decompose a quantity $u$ into a temporally averaged part $u_{\rm mean}$ and a disturbance part $u_{\rm dist}$, such that
\begin{align}
u_{\rm mean}(t) &= \frac{1}{a}\int_{t-a}^{t} u dt^{\prime}, \label{equ:qs_mean}\\
u_{\rm dist}(t) &=~u~-~u_{\rm mean}, \label{equ:qs_dist}
\end{align}
where $a$ represents the width of the time window for averaging.  We carry out the decomposition with $a{=}60$, 180, 300, and 600 seconds. The time integral is performed with a cadence of 100 iterations, which is equivalent to about 10 seconds on average (the timestep for each iteration varies.). We note that the decomposition is based purely on time scales, and the disturbance component may also contain non-wave contributions, such as upsurges of coronal plasma that occur randomly and motions of coronal features (e.g., an expanding coronal loop) that happen to intersect with the slit. For a larger $a$ in the decomposition, more non-wave dynamics of the coronal plasma are classified into the disturbance component. For example, when $a{=}600$\,s, the disturbance component of $I_{193}$ contains all kinds of dynamics seen in the original image, while the mean component only exhibits a slowly evolving background. 

Because the waves seen in \figref{fig:qs_multi} are mostly fast moving disturbances, we present the result of $a=60$\,s to illustrate the short time scale disturbances. As an example, \figref{fig:qs_dist} displays the mean and disturbance of the synthetic AIA 193 intensity ($I_{193}$) and pressure ($p$). An evolution of about 70 minutes starting from $t_{\rm QS}{=}$1\,h\,51\,m\,26\,s is shown by the associated animation, which makes the wave disturbances much clearer for eyes. 

In general, there are two types of propagating disturbances.  One type appears as a single pulse of a large dome-like structure. For example, a jet takes place at $x{\approx}100$\,Mm at the beginning of the animation of \figref{fig:qs_dist} and drives a propagating front. The wave front can be seen at $t_{\rm QS}{\approx}$1\,h\,56\,m in the animation as an arch of intensity and pressure enhancement at $z{\approx}90$\,Mm. The disturbance amplitude is more than 10\% of the mean intensity and pressure. This is more similar to the large-scale waves driven by major solar eruptions, which are also present in the AR run \citep{WangCan+al:2021}. Here, we are more interested in the other type, which is constantly excited from the lower atmosphere and identified as ubiquitous weak enhancement ($<5\%$) in the EUV intensity and pressure. We extract the disturbances of $I_{193}$, $p$, $v_{x}$, and $B_{x}$ for 70 minutes along a slit at $x={113}$\,Mm, yielding the time-distance (T-D) diagrams shown in \figref{fig:qs_timedist}. The slit is placed at a small magnetic flux concentration, above which the magnetic field is dominated by the vertical component. The mean inclination angles of the magnetic field (with respect to the $x-y$ plane) along the slit in the corona below $z{=}20$, 40, and 80\,Mm are about $77^{\circ}$, $71^{\circ}$, and $67^{\circ}$, respectively. Above $z{=}80$\,Mm, the vertical component does not always dominate, and the inclination angle varies more significantly. Nevertheless, the selected slit location offers a relatively simple background for the propagation of waves in the unresting quiet Sun corona.

We can see that the T-D diagrams of $I_{193}$ and $p$ show many straight ridges, corresponding to the upward propagating disturbances along the slit as in the animation of \figref{fig:qs_dist}. The T-D diagrams of $v_{x}$ and $B_{x}$ display more complex features, but the most evident features are curved ridges, indicating a decrease in the propagation speed of the velocity and magnetic field disturbances. To understand these features on the T-D diagrams, we calculate the paths of disturbances traveling with the speed of sound and the Alfv\'en speed by
\begin{align}
P_{cs} &= z_{0} + \int c_{s}dt, \label{equ:qs_pcs} \\
P_{A} &= z_{0} + \int v_{A}dt, \label{equ:qs_pa} 
\end{align}
where $c_{s}$ and $v_{A}$ evolve with time and the integral is performed with a cadence of 100 iterations.
During this time period, the (time-dependent) largest wave speed allowed by the Boris correction is in the range of 700 to 1300\,km/s, and the largest $v_{A}$ unaffected by the Boris correction is less than 900\,km. Moreover, the filling factor of the grid points with $v_{A}>700$ (800)\,km/s is only about 1\% (0.5\%), and these points are mostly found in the atmosphere below 5\,Mm. Therefore, the propagation of Alfv\'en waves in most of coronal volume analyzed here is not affected by the Boris correction. In \figref{fig:qs_timedist}, we plot $P_{cs}$ and $P_{A}$ starting from $z_{0}=1$\,Mm\footnote{To avoid the layers where $c_{s}$ and $v_{A}$ are small.} and 5 time instances that are about 10--15 minutes apart.

The disturbances of $I_{193}$ and $p$ that appear as straight ridges in the T-D diagrams are clearly in line with the paths of sound waves. Moreover, the disturbances of $I_{193}$ and $p$ ($\rho_{\rm dist}$, which is not shown here, is highly similar to $p_{\rm dist}$) coincide in time and position, which can already be seen in \figref{fig:qs_dist}. Thus, the wave mode is compressible and is consistent with the behavior of sound waves. The constant slope of the ridges corresponds to a speed of about 140\,km/s, which matches the speed of sound in the corona.

Similarly, we find that most of the curved ridges in the T-D diagrams of $v_{x}$ and $B_{x}$ disturbances show a good match with the paths of Alfv\'enic disturbances.  We note that the term ``Alfv\'enic disturbances" used here refers to anything that travels with the Alfv\'en speed, without assuming a particular wave mode. As shown by the curved travel path, the Alfv\'en speed is about 500 -- 600 km/s in the low corona below 10\,Mm, gradually decreases to about 150\,km/s at about 40\,Mm height, and becomes about 30\,km/s in the high corona above 100\,Mm. As a side note, the ratio of $c_{s}$ and $v_{A}$ along the slit also reflects the change in plasma $\beta$ with height, as 
$$
\beta = \frac{p}{B^2/(2\mu_{0})} = \frac{2}{\gamma}\frac{c_{s}}{v_{A}} \sim \frac{c_{s}}{v_{A}}.
$$ 

Waves are commonly detected in the lower atmosphere \citep[e.g.,][]{Jess+al:2009,Centeno+al:2009,Grant+al:2018,Schmit+al:2020}, transition region, and corona \citep[see, e.g., ][ and references therein]{DeForest+Gurman:1998,Krijger+al:2001,Tomczyk+al:2007,Srivastava+al:2008,DePontieu+al:2003,DePontieu:2021.IRIS}. \citet{Heggland+al:2011} and \citet{Kato+al:2016} studied the propagation of self-consistently excited waves in 2D simulations that account for the coupling between the solar atmosphere and convection zone. As a comparison, by trading off the high spatial resolution used in the two studies above, the simulation in the paper has a much wider and higher domain, which allows waves to further propagate into the corona. The analysis presented here is mostly for illustration of the presence and basic properties of the waves and does not intend to perform a comprehensive investigation on all possible wave modes, which is beyond the scope of this paper.

\subsection{Energy Fluxes for the Quiet Sun Corona}\label{sec:res_qs_flux}

\begin{figure*}
\includegraphics{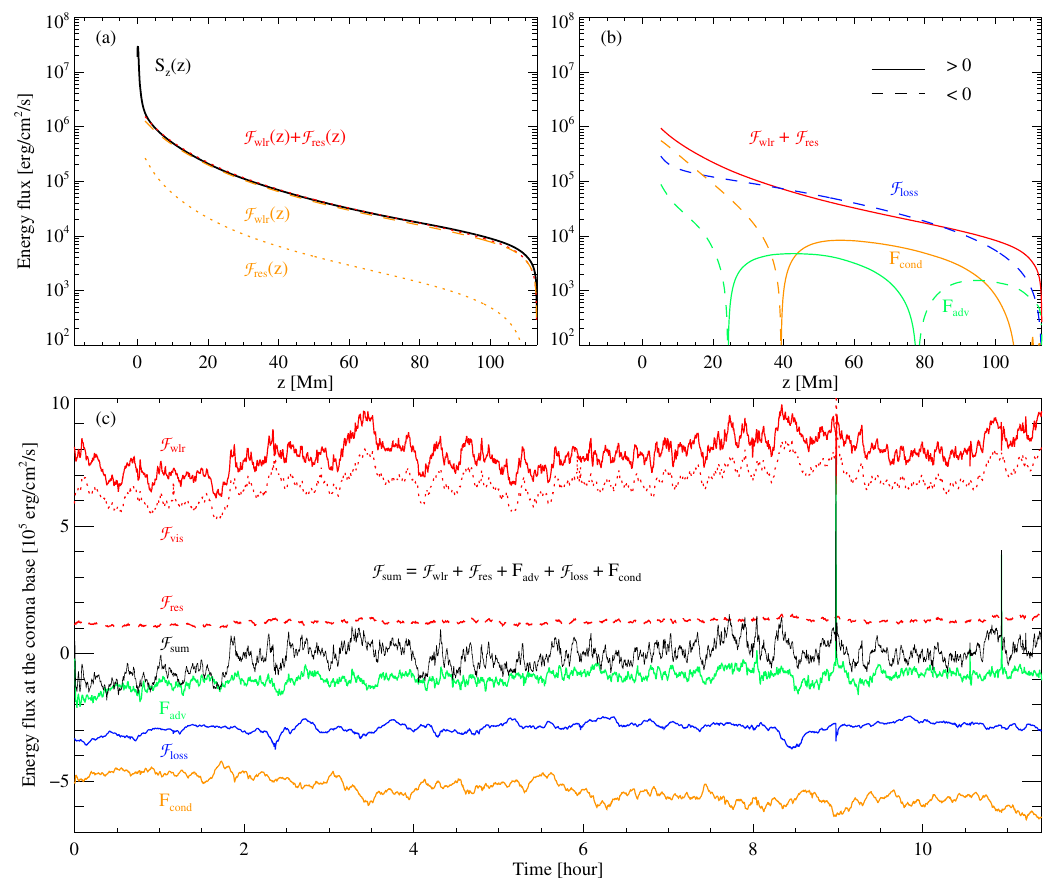}
\caption{Balance of energy fluxes in the quiet Sun corona. (a): Conservation of magnetic energy. The black solid line is the mean vertical Poynting flux ($S_{z}$) on the left-hand side of \equref{equ:eb_cons3}. The red dotted line shows the sum of specific energy fluxes for the Lorentz force work (${\cal F}_{\rm wlr}$) and resistive heating (${\cal F}_{\rm res}$) on the right-hand side of \equref{equ:eb_cons3}. The individual contributions of ${\cal F}_{\rm wlr}$ and ${\cal F}_{\rm res}$ are shown by the orange dashed and dotted lines, respectively.  All these energy fluxes are averaged over 11.5 hours. (b): Balance of temporally averaged energy fluxes on the right-hand side of \equref{equ:ehd}. Lines are noted by the names with corresponding colors. Positive and negative values are shown as solid and dashed lines, respectively. (c): Balance of energy fluxes (also as functions of time) at the base of the quiet Sun corona ($z{=}5.28$\,Mm) as given by \equref{equ:ehd}. Lines are noted by the names with corresponding colors. See \sectref{sec:res_qs_flux} for details.
\label{fig:qs_eflux}} 
\end{figure*}

We investigate the energy fluxes that provide the heating of the quiet Sun corona from the view of the conservation law. The conservation law of magnetic energy reads
\begin{equation}
\label{equ:eb_cons}
\frac{\partial e_{B}}{\partial t} + \nabla\cdot\mathbf{S} = -~Q_{\mathrm{wlr}} - Q_{\mathrm{res}},
\end{equation}
where $e_{B}$ is the magnetic energy density and $\mathbf{S}$ is the Poynting flux. On the right-hand side, $Q_{\mathrm{wlr}}=\mathbf{v}\cdot(\mathbf{j}\times\mathbf{B})$ is the mechanical work done by the Lorentz force, with the current $\mathbf{j}$ given by $\mu_{0}\mathbf{j} = \nabla\times\mathbf{B}$. $Q_{\mathrm{res}}$ is the dissipation that converts magnetic energy into internal energy. In the MURaM code, $Q_{\mathrm{res}}$ is implemented by accurately adding the magnetic energy dissipated by the numerical scheme into the plasma energy.

We integrate \equref{equ:eb_cons} over the volume above a certain height $V(z)$ and investigate the temporal average over the time period of the QS run. Because the quiet Sun in this simulation has reached a dynamic equilibrium, the magnetic energy may fluctuate with time, but its temporal mean over a sufficiently long time period is more or less unchanged. This suggests that the time derivative of the left-hand side vanishes in a statistical sense. The integral of the remainder on the left-hand side yields
\begin{equation}\label{equ:sz_int}
\int \limits_{V(z)} \nabla\cdot\mathbf{S}\,dV = \oint \limits_{\Sigma(z)}\mathbf{S}d\mathbf{\Sigma} = -S_{z}(z) \,L_{x}\,L_{y}, 
\end{equation}
where $\Sigma(z)$ is the surface of $V(z)$ and $S_{z}$ is the mean vertical Poynting flux at height $z$. Because the side boundaries are periodic and vertical fluxes at the top boundary are strongly suppressed, we only need to consider the vertical flux through the horizontal surface at height $z$.

For an energy release rate $Q$ on the right-hand side of \equref{equ:eb_cons}, we can define a specific energy flux ${\cal F}$, such that the integral of $Q$ in this volume is equal to the energy carried by this flux through the surface of the volume. Given the boundary condition discussed above, only the vertical flux through the bottom of the volume matters, which leads to the relation
\begin{equation}\label{equ:f_def}
\int \limits_{V(z)} Q\,dV = {\cal F}(z)\,L_{x}\,L_{y}.
\end{equation}
With the relation in \equref{equ:f_def}, \equref{equ:eb_cons} is reduced to
\begin{equation}
\label{equ:eb_cons3}
S_{z}(z)={\cal F}_{\rm wlr}(z) + {\cal F}_{\rm res}(z),
\end{equation}
where ${\cal F}_{\rm wlr}$ and ${\cal F}_{\rm res}$ are the specific energy fluxes for the Lorentz force work and resistive heating, respectively. This equation describes the conservation of magnetic energy when the 3D domain collapses into the vertical direction.

The vertical profiles of $S_{z}$, ${\cal F}_{\rm wlr}$ and ${\cal F}_{\rm res}$ are averaged over 11.5 hours of evolution and plotted in \figref{fig:qs_eflux}(a). The upward Poynting flux in this simulation is self-consistently generated by the magneto-convection in the convective layer and photosphere and sets the foundation of why a corona of realistic temperature and density can form in the simulation. We also plot ${\cal F}_{\rm wlr}$ and ${\cal F}_{\rm res}$ separately in \figref{fig:qs_eflux}(a). We find that most of the magnetic energy transported by the Poynting flux into the corona is released through Lorentz force work, which is a consequence of using a high numerical magnetic Prandtl number in the simulation. See Section \ref{sec:res_qs_heating} for further discussion.

We also investigate the balance of the energy fluxes that evolve the plasma energy in the corona. When integrated over the volume V(z), the energy equation solved by the simulation,  Equation (30) in \citet{Rempel:2017}, can be reformed as
\begin{equation}\label{equ:ehd}
\frac{\partial E_{\rm HD}}{\partial t}=(F_{\rm adv} + {\cal F}_{\rm wlr} + {\cal F}_{\rm res} + F_{\rm cond} + {\cal F}_{\rm loss})L_{x}L_{y},
\end{equation}
where
$$
E_{\rm HD} = \int \limits_{V(z)}(e_{\rm int} + \frac{1}{2}\rho v^{2}) dV
$$
is the sum of internal energy and kinetic energy in the volume above a given height $z$. Each term on the right-hand side of \equref{equ:ehd} represents a horizontally averaged energy flux at a given height $z$.
$
F_{\rm adv} = (e_{\rm int} + \frac{1}{2}\rho v^{2} + p)v_{z}
$
is the vertical advective flux of enthalpy plus kinetic energy, $F_{\rm cond}$ is the thermal conduction flux, and ${\cal F}_{\rm loss}$ is the specific flux for the energy losses through optically thin radiation. The specific energy flux for the work done by gravity is rather insignificant compared with other physical fluxes ($10^3$ vs. $10^5$ -- $10^6$\,erg/cm$^{2}$/s at the coronal base); thus, it is not shown in this analysis. We also monitor the specific energy flux of the work by the Boris correction ($\mathbf{v} \cdot \mathbf{F}_{\rm SR}$) and the numerical viscous energy flux that accounts for the kinetic energy transported by the numerical viscous stresses (not explicitly written in the equation but added in the numerical code) and find that they are smaller than the physical energy fluxes by two orders of magnitude. Therefore, they are omitted in \equref{equ:ehd}.

Because the quiet Sun corona reaches a dynamic equilibrium and the time derivative of $E_{\rm HD}$ vanishes in a statistical sense, the energy deposition and losses in the corona need to be balanced. Such a balance is demonstrated by the temporal averages of the fluxes on the right hand side of \equref{equ:ehd}, as shown in \figref{fig:qs_eflux}(b). This is similar to the analysis (on heating/cooling rate per mass) shown in \citet{Rempel:2017}, but is now extended to a much larger vertical domain. In the low corona, the total magnetic energy input (${\cal F}_{\rm wlr}+{\cal F}_{\rm res}$) exceeds the energy losses by thermal conduction and radiation. This results in a net energy deposition in the coronal volume. The excessive flux is balanced by a downward advective flux, which has also been found in \citet{Rempel:2017} (see Figure 10 in that study). We also note that the downward advective flux in the low corona is still present even if outgoing fluxes are not damped at the top boundary (done in another experiment not presented here). In the middle domain (e.g., $z{=}60$\,Mm), the radiative loss slightly exceeds the total energy input, as the energy input decays faster than the radiation flux. The difference is mostly balanced by an upward thermal conduction flux with a weaker upward advective flux. The upward conduction flux is not entirely unexpected, because the mean temperature is slightly higher in the low corona as shown in \figref{fig:qs_rhot}. In the higher corona, the energy input prevails again and leads to downward advective flux. Because the magnetic field becomes quite weak and turbulent in the middle and high quiet Sun corona, heat conduction along the magnetic field becomes very inefficient, which was also shown in Figure 11 of \citet{Rempel:2017}. All energy fluxes vanish toward the top boundary, as expected from the boundary condition. At 1\,Mm below the boundary, they are already smaller than the corresponding values at the coronal base by at least two orders of magnitude.

Nevertheless, it makes the most sense to consider the whole coronal volume and evaluate the energy fluxes through the coronal base. \figref{fig:qs_eflux}(c) shows the temporal evolution of the fluxes on the right hand side of \equref{equ:ehd} at the coronal base. In addition, we also show ${\cal F}_{\rm vis}$ in Panel (c), which is the specific energy flux for the viscous heating $Q_{\rm vis}$ in the corona. The viscous heating accounts for the dissipation of kinetic energy that contributes to plasma heating.  $Q_{\rm vis}$ is calculated only for analysis purposes following the same scheme of $Q_{\rm res}$ as described in \citet[][Section 2.6]{Rempel:2017}. This part of energy is {\it implicitly} added to the internal energy because the simulation solves the equation of the plasma energy with a conservative treatment\footnote{When keeping the plasma energy unchanged, a deduction in the kinetic energy contributes to an increase in the internal energy.}. \figref{fig:qs_eflux}(c) demonstrates such a balance that the energy deposition through ${\cal F}_{\rm wlr}$ and ${\cal F}_{\rm res}$ is equalized by the losses through $F_{\rm cond}$, ${\cal F}_{\rm loss}$, and $F_{\rm adv}$. The sum of all fluxes (${\cal F}_{\rm sum}$) fluctuates slightly but is much smaller than the dominating energy deposition (${\cal F}_{\rm wlr}$) and loss ($F_{\rm cond}$) terms through evolution. The temporal average of ${\cal F}_{\rm sum}$ is about $7\times10^{3}$\,erg/cm$^{2}$/s, which is less than 1\% of ${\cal F}_{\rm wlr}$. It is interesting to note that two eruptions that are even evident in the volume integrated energy occur in the QS run at $t{\approx}9$\,h and 11\,h, respectively. They are very short transients and do not affect the temporally averaged quantities.

To summarize, the upward Poynting flux $S_{z}{=}9.03\times10^{5}$\,erg/cm$^{2}$/s provides the magnetic energy input and is balanced by ${\cal F}_{\rm wlr}{=}7.90\times10^{5}$\,erg/cm$^{2}$/s and ${\cal F}_{\rm res}{=}1.25\times10^{5}$\,erg/cm$^{2}$/s, which release magnetic energy to plasma energy. ${\cal F}_{\rm wlr}$ is deposited first in the kinetic energy and then quickly converted to plasma heating, as ${\cal F}_{\rm wlr}$ is mostly balanced with ${\cal F}_{\rm vis}{=}6.75\times10^{5}$\,erg/cm$^{2}$/s, and their temporal evolution is highly similar. The magnetic energy flux into the corona contributes a direct heating of ${\cal F}_{\rm vis}+{\cal F}_{\rm res}{=}8\times10^{5}$\,erg/cm$^{2}$/s to coronal plasma. The residual of ${\cal F}_{\rm wlr}-{\cal F}_{\rm vis}{=}1.15\times10^{5}$\,erg/cm$^{2}$/s is balanced by the work against pressure forces and results in the downward directed enthalpy flux.

\subsection{Implications on Coronal Heating}\label{sec:res_qs_heating}
This simulation successfully produces a stable and self-maintained million K hot corona, which is a fundamental aspect of the coronal heating problem \citep{Klimchuk:2006.review,Klimchuk:2015.review}. We note that the heating of the quiet Sun in this simulation is essentially identical to the Quiet Sun and Coronal Arcade simulations in \citet{Rempel:2017}, which presented a thorough analysis of the coronal heating process. Therefore, we do not repeat the analysis that has been done in that study, and the analysis presented in this paper yields consistent results in comparison to the simulations in \citet{Rempel:2017}. We briefly discuss the following questions related to coronal heating.

Despite the debate on whether the energy for coronal heating is provided by waves \citep{Kudoh+Shibata:1999,Moriyasu+al:2004,vanBallegooijen+al:2011,Arregui:2015} or fieldline braiding \citep{Parker:1983,Galsgaard+Nordlund:1996,Priest+al:2002,Gudiksen+Nordlund:2005a,Rappazzo+al:2008,Wilmot:2015}, it is generally accepted that in both scenarios, the energy is generated by the interaction of the magnetic field and granular motions in the photosphere. In this simulation, we do not presume on what time and spatial scales this interaction happens, instead, an upward energy flux ($S_{z}$) of about $3\times10^{7}$\,erg/cm$^{2}$/s in the photosphere, which is similar to observed values \citep[e.g.,][]{Yeates+al:2014,Welsch:2015}, is self-consistently generated by magneto-convection. $S_{z}$ drops to about $9\times10^{5}$\,erg/cm${^2}$/s at $z{=}5.28$\,Mm, which provides a sufficient energy budget for the heating of the quiet Sun corona compared with the requirement estimated from observations \citep{Withbroe+Noyes:1977}. 

As shown in \figref{fig:qs_eflux}(c), the input energy fluxes are a composition of a relatively steady background and a fast varying fluctuation component that are supposed to represent the braiding and wave contributions, respectively. \citet{Rempel:2017} analyzed the contribution to the Poynting flux into the corona by dynamics on different time scales and showed that the Poynting flux distributes smoothly over a wide range of time scales and that there is no hard separation between the contributions from waves and braiding. Furthermore, the energy flux into the corona above 4.8\,Mm (similar height to the coronal base defined in this study) is mainly generated by the velocity and magnetic field on time scales larger than 10 minutes (to be compared with the Alfv\'en time scale of 2.5 minutes), which clearly suggests that the energy flux is dominated by slow braiding of the magnetic field.

In addition to direct energy deposition in the corona, heating events deep down in the lower atmosphere can also generate upflows of heated plasma that feed into the corona \citep[e.g.,][]{DePontieu+al:2011,Samanta+al:2019}. The jets and surges shown in \figref{fig:qs_multi} indeed carry plasma of coronal temperatures, and their contribution can be measured as enhanced upward enthalpy flux. However, the downward directed average enthalpy flux at the coronal base demonstrates that upflows of heated plasma associated with these jets and surges are not a net source of hot plasma in the corona (at least not in this simulation). The same assessment is also applicable to coronal bright points and campfire-like events. These heating episodes generally occur at lower heights \citep[see also][]{ChenYajie+al:2021} and can indeed heat local plasma. However, they do not significantly impact the large-scale corona above. The dominating energy deposition is the dissipation of the Poynting flux in the extensive corona.

The partition of the magnetic energy release and plasma heating shown in \figref{fig:qs_eflux} indicates that the Lorentz force work is much more efficient than resistive heating for releasing magnetic energy, and consequently, the viscous dissipation contributes on average five times more than the resistive dissipation to the heating of the coronal plasma. This is determined by the high magnetic Prandtl number of this simulation \citep{Rempel:2017,Brandenburg+Rempel:2019}, which is actually a reasonable representation of the situation of the real corona. As estimated by the ratio of the viscosity and resistivity given by \citet{Spitzer:1962}, the corona has an extremely high magnetic Prandtl number ($10^{10}$). \citet{Rempel:2017} showed that the ratio of viscous and resistive heating strongly depends on the magnetic Prandtl number; however, the sum of both are insensitive to the magnetic Prandtl number. The latter is a reflection of energy conservation, which means that in a quiescent corona the total heating (plus the partition into the kinetic energy, if any) is determined by the Poynting flux that provides the available energy budget. 

Furthermore, energy conservation also implies that whether the corona can be heated to a sufficiently high temperature is not determined by the dissipation coefficients used in a model. Even though the dissipation coefficients (particularly the resistivity in those models based on Ohmic heating) used in large-scale 3D simulations \citep{Gudiksen+Nordlund:2005a,Gudiksen+Nordlund:2005b,Bingert+Peter:2011,Bourdin+al:2013,Chen+al:2014,Hansteen+al:2010,Hansteen+al:2015} are supposed to be much larger than the values in the real corona, the reason why the corona or coronal loops in these models can be successfully heated to high temperatures is a high upward Poynting flux that is found either over the whole coronal base or at the footpoints of coronal loops.

The numerical scheme of the MURaM code uses numerical diffusivities that are highly non-uniform in space. Nevertheless, \citet{Rempel:2017} estimated that the effective viscosity and resistivity are on the order of $10^{13}$ and $10^{11}$\,cm$^{2}$/s, respectively. Although the numerical resistivity is still significantly larger than the resistivity given by \citet{Spitzer:1962} for typical coronal conditions, the viscosity in this simulation is on average similar to the Spitzer value of $10^{14}$\,cm$^{2}$/s.\footnote{The viscosity can also be anisotropic in the presence of a magnetic field \citep{Braginskii:1965,MacTaggart+al:2017}, which is not considered in the present simulation.} The conversion of magnetic energy into internal energy is primarily the viscous dissipation of the flows driven by the Lorentz force on the grid-spacing scale. The role of the resistivity is to allow for topology changes through reconnection, but direct resistive heating is small. We suggest this is also likely the process in effect in the high magnetic Prandtl Number corona. This scenario of energy conversion is similar to that in reconnection models \citep{Petschek:1964,Longcope:2009}, although the resolution used in our simulation is not sufficient to resolve microscopic processes of magnetic reconnection.

\section{RESULTS II: EMERGENCE OF ACTIVE REGIONS FROM THE CONVECTION ZONE TO THE CORONA}\label{sec:res_ar}
\subsection{Formation of Active Regions in the Photosphere}\label{sec:res_ar_form}

\begin{figure*}
\center
\includegraphics{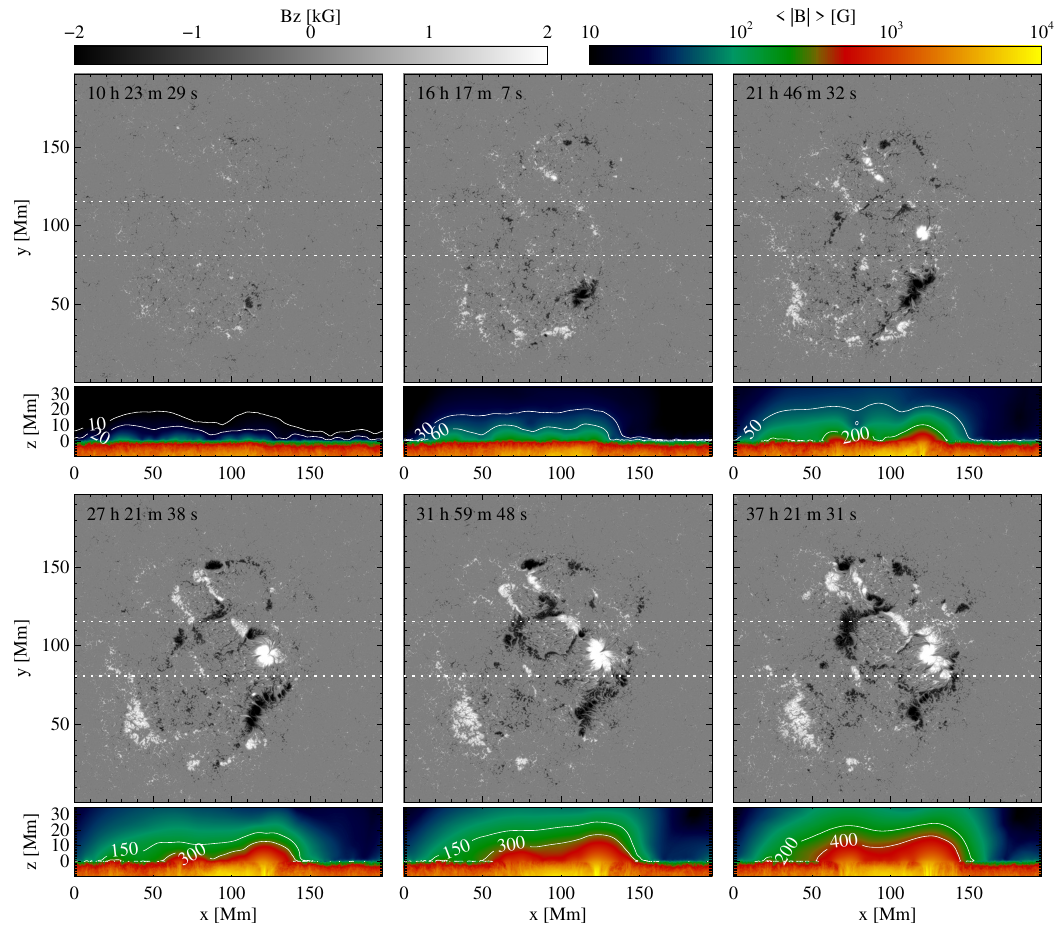}
\caption{Evolution of the magnetic field in the photosphere and low corona as active regions emerge. Grayscale images in square panels present $B_{z}$ at the $\tau{=}1$ surface at six time instances. The color-scale panel below each grayscale image shows the magnetic field strength in the first 45\,Mm of the vertical domain (from 10\,Mm beneath to 35\,Mm above the photosphere), overlaid by two contour lines that illustrate the asymmetry of coronal magnetic field strength. The magnetic field strength is averaged in the $y$ direction between the two dotted lines in the $B_{z}$ maps.
\label{fig:ar_btau_vert}} 
\end{figure*}

\begin{figure*}
\center
\includegraphics{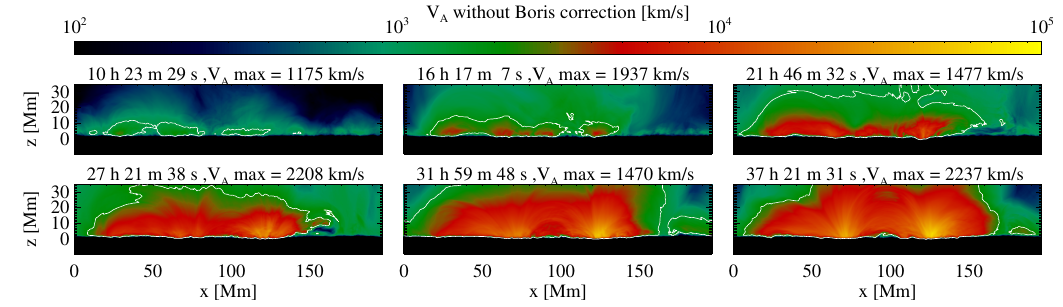}
\caption{Mean (unlimited) Alfv\'en velocity in the first 45\,Mm of the vertical domain (from 10\,Mm beneath to 35\,Mm above the photosphere). The Alfv\'en velocity is averaged in the $y$ direction between the two dotted lines in \figref{fig:ar_btau_vert}. Time instances are identical to those in \figref{fig:ar_btau_vert}. The actual maximum Alfv\'en velocity allowed by the Boris correction is marked above each panel and plotted as a contour line on the map of the mean Alfv\'en velocity. 
\label{fig:ar_va_vert}} 
\end{figure*}

We use the QS run snapshot at $t_{QS}{=}5.7$\,h as the initial snapshot for the AR run after cutting out the lowermost 8\,Mm of the domain and switching on the data-driven boundary condition that imposes the velocity and magnetic fields adapted from the solar convective dynamo simulation. The evolution of the imposed magnetic field in the uppermost convection zone and the formation of active regions in the photosphere have been presented in detail in Paper\,I.

Although the depth of the convective layer is slightly deeper, the evolution of the photosphere and convection zone in the AR run is basically the same as the ``$196\times8$ run" in Paper\,I. A very brief recap of the active region evolution is presented by the magnetograms ($B_{z}$ at the photosphere) in \figref{fig:ar_btau_vert}. Two major flux bundles reach the photosphere and form the largest two sunspot pairs. The first flux bundle reaching the photosphere appears at $t{\approx}10.5$\,h in the lower half of the horizontal domain. After $t{\approx}16$\,h, the magnetic field related to the strongest flux bundle emerges in the photosphere near the y-center of the domain and develops into a strong sunspot at $t{\approx}22$\,h. Over the whole evolution of about 50 hours, a few other flux bundles also emerge into the photosphere and create smaller sunspot pairs, for example, at $y{\approx}150$\,Mm. We also note that convection can destroy the structure of a coherent flux bundle during the course of emergence. The magnetic field removed from a strong flux bundle can be transported to the surface by convective upflows. This process gives rise to many magnetic concentrations of about 10\,Mm in diameter. The small sunspot pairs and magnetic concentrations add a significant complexity to the magnetic field in the simulation and play a crucial role in producing solar eruptions in the active regions.

The small panel below each magnetogram in \figref{fig:ar_btau_vert} displays the development of the strong sunspot pair in the center of the domain. The quantity shown here is the magnetic field strength averaged between $y{\approx80}$ and 115\,Mm, i.e., between the two dotted white lines drawn on the magnetograms. The 45\,Mm vertical field-of-view of the small panels spans from 10\,Mm beneath to about 35\,Mm above the photosphere. The higher domain is not shown here. Two contours are drawn in each panel to illustrate the asymmetry in the magnetic field strength in the corona above the asymmetric photospheric sunspots.

In the first 10 hours when this flux bundle has not yet entered the domain through the bottom boundary, the convection zone is filled with small-scale magnetic flux maintained by a small-scale dynamo (also known as local dynamo). When the flux bundle emerges close to the photosphere and gives rise to a few small and fragmented flux concentrations, it already leads to a significant increase in the mean coronal magnetic field strength from about 10\,G at $t{\approx}10.5$\,h to over 60\,G at $t{\approx}16$\,h. The flux concentration of the leading polarity, which is located at $(x,y){\approx}(120,95)$, becomes a strong and coherent sunspot at $t{\approx}22$\,h. At this moment, we can see that the strong vertical magnetic flux tube beneath the photospheric sunspot significantly expands in the corona and creates a dome-like area where the mean field strength is about 300\,G. After the sunspot pair becomes fully developed, the mean field strength in the lowermost 35\,Mm above the photosphere can reach about 500\,G.

The sunspot pairs formed in this simulation show clear asymmetries that are similar to those found in observed sunspots, in the sense of a stronger and more coherent leading spot and weaker and less coherent trailing spot. These properties are caused by the asymmetric flows associated with the flux bundles generated in the solar convective dynamo as explained in Paper\,I. It is not surprising that the asymmetry in magnetic field strength extends to, if not the whole vertical domain, the lower corona. The coronal magnetic field strength on the leading spot side is clearly stronger than that on the trailing spot side. The difference is illustrated in the small panels by the contour lines that are inclined between the two sunspots. At the three time instances between $t{\approx}22$ and 32\,h, when the asymmetry is the most evident, the mean magnetic field strength (averaged between $z{=}0$ and 35\,Mm) above the leading sunspot (about 230, 450, 600\,G) is about two times larger than that above the trailing sunspot (about 130, 200, 300\,G). Later, when the trailing sunspot becomes stronger, the asymmetry becomes less significant but is still present.

To demonstrate the necessity of using the Boris correction in active region simulations, we present in \figref{fig:ar_va_vert} the mean unaffected Alfv\'en velocity in the first 45\,Mm height in the central region of the domain. The chosen region and time instances are the same as for evaluating the mean magnetic field strength shown in \figref{fig:ar_btau_vert}. With the development of strong sunspots in the photosphere, the largest values of the mean Alfv\'en velocity grow from around 1000\,km/s in the early stage of magnetic flux emergence ($t{\approx}10.5$\,h) to about $1.8\times10^{4}$\,km/s after the first coherent sunspot is formed ($t{\approx}22$\,h) and to about $6.8\times10^{4}$\,km/s when sunspots are fully formed ($t{\approx}37.5$\,h). Meanwhile, the filling factor of the high Alfv\'en velocity region also increases with time. At $t{\approx}37.5$\,h, the 35\,Mm vertical domain above the photosphere has $V_{A}{>}5000$\,km/s on average. We also mark the actual maximum Alfv\'en velocity as limited by the Boris correction ($V_{A~\mathrm{max}}$) above each panel in \figref{fig:ar_va_vert}. The peak (not averaged) unlimited Alfv\'en velocity in the same region is $10^4$\,km/s (in the early stage) and $2\times10^5$\,km/s (in the late stage). Therefore, the Boris correction provides an improvement of at least 10 to 100 times in timesteps, with negligible impacts on the dynamics of the system.

We plot in \figref{fig:ar_mag_energy}(a) the total unsigned magnetic flux at the photosphere, as well as the flux emergence rate, which is given by the time derivative of the magnetic flux. The total unsigned magnetic flux for the 197$^{2}$\,Mm$^{2}$ quiet Sun is $2\times10^{22}$\,Mx. After 10 hours, as emerging flux bundles arrive in the photosphere, the magnetic flux begins to increase at a relatively stable rate of about $4\times10^{21}$\,Mx/h. The flux continues to grow to about $1.3\times10^{23}$\,Mx at the end of the AR run. In the latter half of the evolution, the mean emergence rate also decreases to about $2\times10^{21}$\,Mx/h. The evolution from $t{=}8$ to 48\,hours leads to a mean flux emergence rate of $2.5\times10^{21}$\,Mx/h. 

Considering the domain as a whole active region, its magnetic flux is comparable to the largest active regions observed on the Sun, for example, NOAA active region 12192, which contains $1.6\times10^{23}$\,Mx unsigned flux \citep{SunXudong+al:2015}. This large simulated active region is accompanied by a faster flux emergence than smaller active regions \citep{Otsuji+al:2011,Toriumi+al:2014,Norton+al:2017}\footnote{The unsigned flux and emergence rate shown in \figref{fig:ar_mag_energy}(a) are divided by 2 when they are compared with the signed flux and emergence rate measured within a single polarity \citep[e.g., as in][]{Norton+al:2017}.}. \citet{SunXudong+Norton:2017} reported the fastest flux emergence ever observed in an active region with about half of the peak flux of this simulation. The observed mean (peak) emergence rate is about $4.9\times10^{20}$ ($1.1\times10^{21}$) Mx/h, which is about $1/5$ ($1/4$) of the emergence rate in the simulation. The numerical experiments in Paper\,I showed that when the bottom boundary is placed deeper (e.g., 18\,Mm and 30\,Mm beneath the photosphere), it takes longer for flux emergence and less unsigned flux emerges to the photosphere, yielding a reduced flux emergence rate (about $1/3$ of the current simulation) that is closer to observations.

\subsection{Evolution and Distribution of Magnetic Energy}\label{sec:res_ar_mag}

\begin{figure*}
\center
\includegraphics[width=1.0\textwidth]{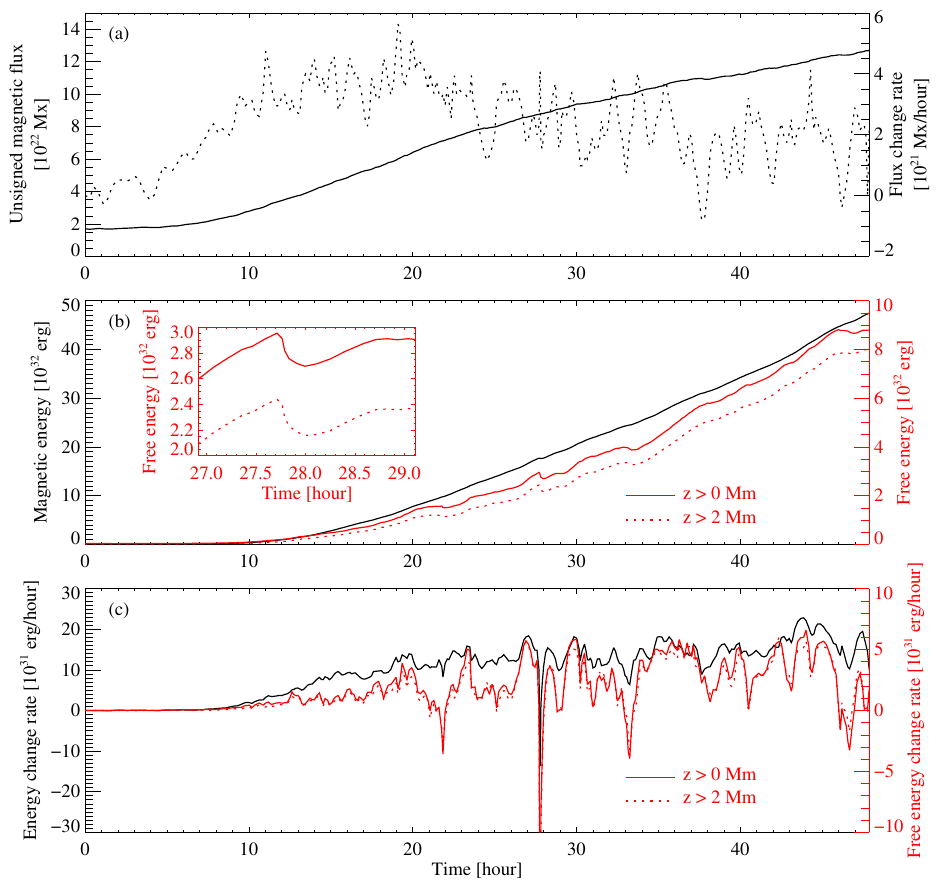}
\caption{Temporal evolution of magnetic flux that emerged in the photosphere and the magnetic energy contained in the domain above the photoshere. (a): The solid line shows the total unsigned vertical magnetic flux. The dotted line is the flux emergence rate given by the time derivative of the magnetic flux. (b): The black and red lines plot the total magnetic energy ($E_{\rm tot}$) and free magnetic energy ($E_{\rm free}$), respectively. In particular, the solid red line is for $E_{\rm free}$ in the domain above the photosphere, and the dotted red line only accounts for $E_{\rm free}$ in the domain above $z{=}2$\,Mm, which corresponds to the low $\beta$ corona. A zoomed view of a time period of two hours showing the evolution of the free magnetic energy during an eruption event is presented in the insert. (c): Rate of change of the total and free magnetic energy from panel (b).
\label{fig:ar_mag_energy}} 
\end{figure*}

\begin{figure*}
\center
\includegraphics[width=1.0\textwidth]{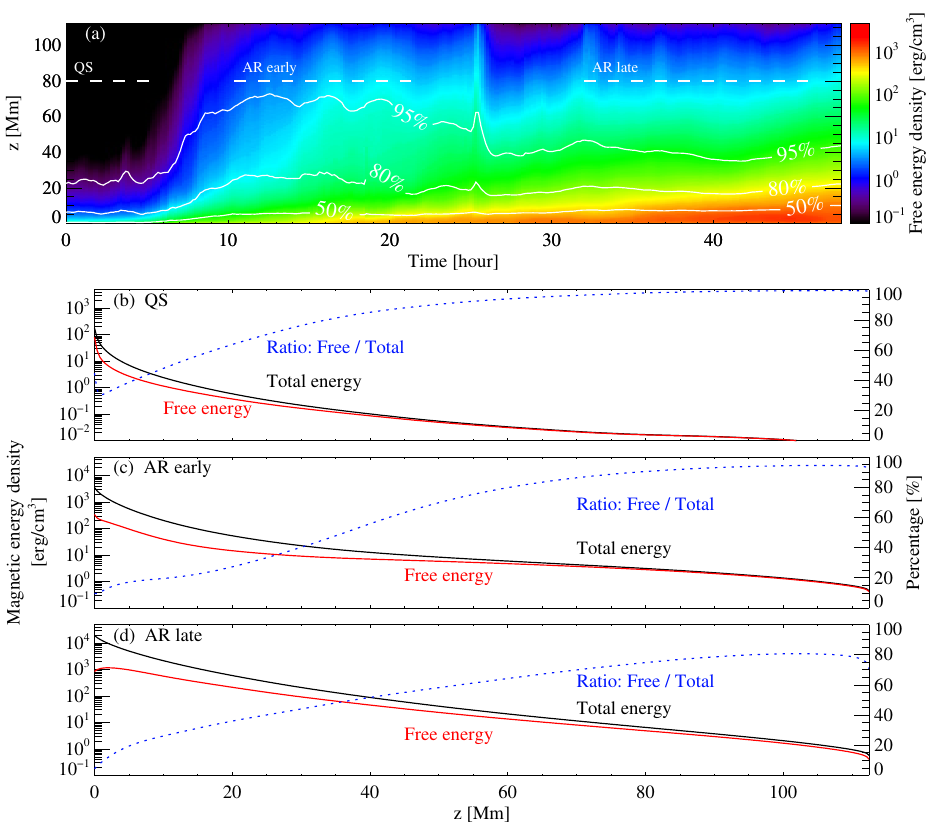}
\caption{Distribution of the free magnetic energy with height ($z{>}0$\,Mm). (a) Time-height diagram showing the evolution of the horizontally averaged free magnetic energy density ($e_{\rm free}$). The contour lines of $f_{\rm cum}$ indicate the percentage of the free energy contained in the volume below a certain height, as given by \equref{equ:ecum}. The dashed lines mark three evolution stages, over which temporal averages of $e_{\rm free}$ are calculated. (b): Horizontally and temporally averaged magnetic energy density in the quiet Sun (QS). The solid black line plots the total magnetic energy density ($e_{\rm tot}$), and the solid red line is $e_{\rm free}$. The dotted black line is the ratio between the red and black lines, i.e., the percentage of the free energy among the total magnetic energy. (c): The same analysis as in Panel (b) for the early stage of active region evolution (AR early). The notations of the lines are identical to those in Panel (b). (d): The same analysis as in Panel (b) for the late stage of active region evolution (AR late). The notations of the lines are identical to those in Panel (b).
\label{fig:ar_mag_energy_z}}
\end{figure*}

In this section, we study the temporal evolution and spatial distribution of the magnetic energy in the atmosphere.

The total magnetic energy in the atmosphere, $E_{\mathrm{tot}}$ is given by
\begin{equation}\label{equ:ebt}
E_{\rm tot} = \int\limits_{V(z{>}0)} \frac{\mathbf{B}^{2}}{2\mu_{0}}\,dV.
\end{equation}
Then, we obtain the free magnetic energy, $E_{\mathrm{free}}$, by 
\begin{equation}\label{equ:ebf}
E_{\rm free} = \int\limits_{V(z{>}0)} \frac{\mathbf{B}^{2}-\mathbf{B}^2_{\rm pot}}{2\mu_{0}}\,dV,
\end{equation}
where $\mathbf{B}_{\rm pot}$ is the potential magnetic field computed from $B_{z}$ at $z{=}0$\,Mm.  Calculation of $\mathbf{B}_{\rm pot}$ uses the condition that horizontal boundaries are periodic and $B_{z}$ vanishes at infinity because the signed vertical magnetic flux in any horizontal layer is balanced in this simulation. The temporal evolution of the magnetic energy is shown in \figref{fig:ar_mag_energy}(b). In particular, we plot two lines for the free magnetic energy. The solid line is for the domain above the photosphere, as defined in \equref{equ:ebf}. The dotted line only accounts for the domain of $z{>}\,2$Mm \footnote{using the same potential field computed from $B_{z}$ at $z{=}0$\,Mm} when evaluating the integral in \equref{equ:ebf}. This part of the domain belongs to the low plasma $\beta$ regime, as shown later in this paper in \sectref{sec:res_ar_force}. The change rates of the total and free magnetic energy are evaluated by their time derivatives and are plotted in \figref{fig:ar_mag_energy}(c) with the same line style as in Panel (b).

The quiet Sun atmosphere contains $8.5\times10^{30}$\,erg of $E_{\rm tot}$ in the {197}$^{2}\times{113}$\,Mm$^{3}$ volume. During the first few hours when the emerging flux bundles are mostly in the convection zone, $E_{\rm tot}$ increases very slowly to $3.5\times10^{31}$\,erg at about $t{=}10$\,h. As the active region magnetic field reaches the photosphere and further expands into the corona, $E_{\rm tot}$ starts to grow rapidly and is increased to $7.8\times10^{32}$\,erg at $t{=}20$\,h, with an average growth rate of $7.4\times10^{31}\,$erg/h. Thereafter, $E_{\rm tot}$ grows at a relatively stable rate of $1.38\times10^{32}\,$erg/h and reaches about $4.7\times10^{33}$\,erg by the end of the simulation. 

The curve of $E_{\rm free}$ follows the general trend of $E_{\rm tot}$, as shown in \figref{fig:ar_mag_energy}(b), and their overall temporal evolution is clearly correlated with the photospheric magnetic flux (linear correlation coefficient $> 0.95$). This is a natural consequence of the fact that the pre-existing magnetic field in the quiet Sun in this simulation is weak, and the magnetic energy in the atmosphere is dominated by the newly emerged strong active region magnetic field.

The similarity between the curves of $E_{\rm free}$ and $E_{\rm tot}$ implies that the percentage of the free energy, i.e., the ratio of $E_{\rm free}$ and $E_{\rm tot}$, is a relatively constant value during evolution. In the time period between $t{=}10$ and 48\,h, $E_{\rm free}/E_{\rm tot}$ varies between 15\% and 20\% with a mean value of 17.9\%, which is equivalent to $E_{\rm tot}/E_{\rm pot}=1.22$, where $E_{\rm pot}$ is the energy of the potential field that has been used to evaluate $E_{\rm free}$ in \equref{equ:ebf}. In investigations of observed solar active regions, it seems that there is no general agreement on the percentage of free energy in the total magnetic energy. In addition to the fact that this ratio is likely to be highly case-dependent, the estimate of the magnetic energy in a particular active region relies on force-free modeling of the coronal magnetic field \citep{Wiegelmann+Sakurai:2012,Wiegelmann+al:2014}. The magnetic energy estimated by different extrapolation methods is noticeably different, and the values of $E_{\rm tot}/E_{\rm pot}$ are usually between 1.0 and 1.3 \citep{Schrijver+al:2008,DeRosa+al:2009}. \citet{Aschwanden+al:2014} obtained $E_{\rm tot}/E_{\rm pot}{=}1.106$ from 172 samples, which means $E_{\rm free}/E_{\rm tot}{\approx}9.6\%$.

During certain times, $E_{\rm free}$ shows a sudden decrease. One such case is highlighted by the insert in \figref{fig:ar_mag_energy}(b), where $E_{\rm free}$ drops by more than $2\times 10^{31}$\,erg in about 20 minutes with a peak energy decay rate exceeding $10^{32}$\,erg/h. These sudden decreases are caused by eruptions (flares and coronal mass ejections) occurring in the simulation. The events already visible in the $E_{\rm free}$ curve (note that this is integrated over the whole domain) are very energetic ones with a decrease in magnetic energy on the order of $10^{31}$ erg. Many smaller events are present as well. They are hardly discerned in \figref{fig:ar_mag_energy}(b) but are clearly illustrated by the change rate of $E_{\rm free}$ shown in \figref{fig:ar_mag_energy}(c). The behavior of gradual build up and sudden releases of magnetic (free) energy is highly consistent with the observations of emerging active regions that lead to solar eruptions, for example, NOAA active regions 11072 \citep{LiuYang+Schuck:2012}, 11158 \citep{SunXudong+al:2012,LiuYang+Schuck:2012,Tziotziou+al:2013}, 11283 \citep{JiangChaowei+al:2014,JiangChaowei+al:2016}, and 11817 \citep{LiuRui+al:2016}. 

In addition to the temporal evolution of the magnetic energy in the whole domain, it is also interesting to investigate the distribution of magnetic energy, particularly the free energy, within the domain. For this purpose, we evaluate the horizontally averaged free energy density, $e_{\mathrm{free}}(z)$, by
\begin{equation}\label{equ:ebf_z}
e_{\rm free}(z) = \frac{1}{L_{x}L_{y}}\int \limits_{0}^{L_{y}} \int \limits_{0}^{L_{x}} \frac{\mathbf{B}^{2}-\mathbf{B}_{\rm pot}^{2}}{2\mu_{0}}\,dx\,dy.
\end{equation}
The horizontally averaged total magnetic energy, $e_{\rm tot}(z)$, can be estimated in the same manner. Furthermore, we evaluate the following factor, $f_{\rm cum}$, by
\begin{equation}\label{equ:ecum}
f_{\rm cum}(z) = \frac{\int_{0}^{z} e_{\rm free}(z^{\prime})dz^{\prime}}{\int_{0}^{L_{z}} e_{\rm free}(z^{\prime})dz^{\prime}},
\end{equation}
which describes how much of the free energy is contained in the volume below a certain height, and obviously, it equals unity at $z=L_{z}$. \figref{fig:ar_mag_energy_z}(a) shows a time-height diagram that illustrates the distribution of $e_{\rm free}$ through the vertical domain and evolution through the whole course of the simulation, with the contour lines indicating $f_{\rm cum}$ levels of 50\%, 80\% and 95\%.

A quantitative comparison between the distribution of free magnetic energy above the quiet Sun and active regions (in early and later evolution stages) is shown in \figref{fig:ar_mag_energy_z}(b), (c), \& (d). For the quiet Sun stage, we calculate the temporal average of $e_{\rm free}$ and $e_{\rm tot}$ over a time period of 5.6 hours starting from $t{=}0$\,h, and for the active region, the temporal average is calculated over a time period of 11.4 (14.4) hours starting from $t{\approx}10 (32)$\,h for the early (late) evolution stage, as marked by the dashed lines in \figref{fig:ar_mag_energy_z}(a).

Before the emerging magnetic flux reaches the photosphere, the atmosphere is still dominated by the small-scale magnetic field in the quiet Sun. The contour lines in \figref{fig:ar_mag_energy_z}(a) suggest that 50\% of the free energy is concentrated in the volume below $z{=}1.2$\,Mm and that the lowermost 20\,Mm above the photosphere already contains most ($\sim95\%$) of the free energy. \figref{fig:ar_mag_energy_z}(b) shows that in the quiet Sun, both $e_{\rm free}$ and $e_{\rm tot}$ drop quickly with height in the first few Mm and continue to decrease on a flatter slope (on a logarithmic scale). Because $e_{\rm tot}$ decays slightly faster than $e_{\rm free}$, they are almost equal above $z{\approx}50$\,Mm. Their ratio (blue dotted line) also indicates that the upper half of the domain is dominated by free energy (close to 100\%). This is because the (upper) quiet Sun corona is in the high $\beta$ regime, as mentioned in \sectref{sec:res_qs_wave}, and the flow field gives rise to a turbulent and highly non-potential magnetic field. If we integrate for the energy at all heights (i.e., $E_{\rm tot}$ and $E_{\rm tot}$ shown in \figref{fig:ar_mag_energy}), the free energy is about 40\% of the total magnetic energy in the quiet Sun atmosphere.

The ``AR early'' stage represents a dynamic transition stage from a pure quiet Sun to a developed active region through fast flux emergence. \figref{fig:ar_mag_energy_z}(c) shows that, as the emerged active region magnetic field quickly fills the domain, the magnetic energy in the upper half of the corona is increased by more than 100 times from the quiet Sun values, which makes $e_{\rm free}$ above 50\,Mm about comparable to that above well developed active regions in the late stage of evolution. This stage exhibits the most extensive distribution of the free energy with height, as illustrated by the rather flat slope of $e_{\rm free}$ ($e_{\rm tot}$ as well) above $z{=}40$\,Mm. Meanwhile, in the lower half of the domain, the increase in $e_{\rm tot}$ is clearly more than the increase in $e_{\rm free}$, which is also demonstrated by the large decrease of their ratio (e.g., from 60\% to 20\% at $z{=}20$\,Mm and from 30\% to less than 10\% near the photosphere). This suggests a significant growth in the potential component, as strong and stable magnetic flux concentrations form in the photosphere.

During the ``AR late" stage flux, emergence is still ongoing, and coronal magnetic energy continues to grow. \figref{fig:ar_mag_energy_z}(d) shows that both $e_{\rm free}$ and $e_{\rm tot}$ in the lower half of the domain increase significantly by an order of magnitude, and their vertical profiles decay exponentially (almost constant slope on logarithmic scale) for most of the atmosphere. The contour lines in Panel (a) show that about 50\%\,(95\%) of the free energy is contained in the first 10\,(40)\,Mm, which is a much greater height range compared with the quiet Sun situation shown in Panel (b). The percentage of free energy in the total magnetic energy  increases with height in most of the vertical domain from about 5\% near the photosphere to about 80\% above 100\,Mm. It is very interesting to note that even above strong active regions, the magnetic field in the high corona could be highly non-potential. When integrated over the whole atmosphere, the free energy is about 18\% of the total magnetic energy.

\subsection{Force-freeness of the Magnetic Field}\label{sec:res_ar_force} 
\begin{figure*}
\center
\includegraphics{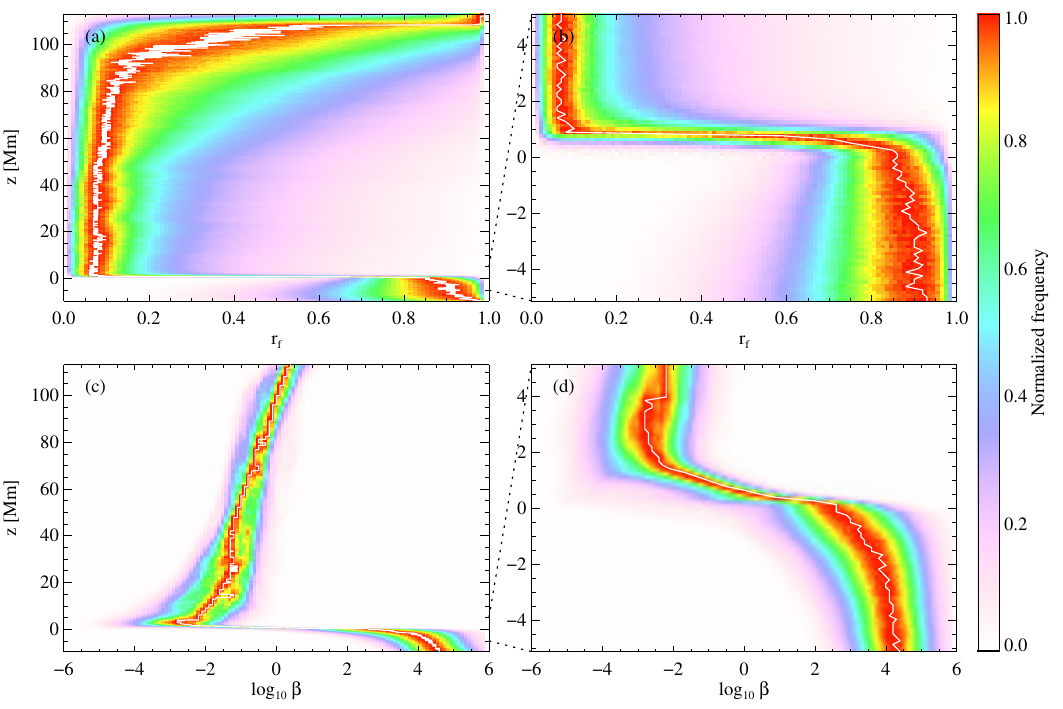}
\caption{Transition from the strongly non-force-free magnetic field and high $\beta$ plasma in the interior to the (close to) force-free magnetic field and low $\beta$ plasma in the corona. (a) The probability density function (PDF) of $r_{f}$ defined by \equref{equ:fff}. (b) A zoomed view of the same quantity as in Panel (a) in the domain between $z{=}-5$ and 5\,Mm. (c) The PDF of plasma $\beta$. (d) A zoomed view of the same quantity in Panel (c) in the domain between $z{=}-5$ and 5\,Mm. The white line in each panel indicates the most probable value of the quantity shown in that panel.
\label{fig:ar_forcebeta}} 
\end{figure*}

It has been commonly accepted that magnetic energy dominates over plasma energy in the corona, which also implies that plasma $\beta$, the ratio of gas pressure to magnetic pressure, is much smaller than unity. In a static low $\beta$ corona, if the gas pressure is ignored, the Lorentz force in the magnetic field must vanish. This leads to the widely exploited force-free modeling of the coronal magnetic field \citep{Wiegelmann+Sakurai:2012,Wiegelmann+al:2014}.

In a realistic model, the non-vanishing gas pressure gradient, together with generally all other non-vanishing terms in the momentum equation, must be balanced by the Lorentz force, which leads to deviations from the force-free state. The corona is heated by dynamical processes that are a combination of slow braiding of field lines (as in DC heating) and waves (as in AC heating). We have shown in \sectref{sec:res_qs} that the magnetic energy released through the Lorentz force work is balanced by viscous heating, suggesting a balance between the Lorentz force and viscous stress. Moreover, in a dynamical corona, the evolution of the magnetic field is driven by the Lorentz force, which can be significant. This raises the question of how force-free the coronal magnetic field is and how reliable the modeling of coronal magnetic field is \citep{Peter+al:2015}. Moreover, the first few Mm above the photosphere are a region where the magnetic field and plasma thermodynamics drastically change, which certainly leads to a great change in $\beta$ and force-freeness of the magnetic field.

The simulation presented in this study is an exceptional test case for evaluating magnetic forces with a non-vanishing pressure gradient force of the plasma. We summarize a few reasons. First, its domain covers the two extreme regimes, namely, the plasma-dominating interior and magnetic-dominating corona. Then, by including realistic physical processes in the energy transport, the simulation produces realistic thermodynamic quantities, which can provide a more sensible representation of the plasma forces in the solar atmosphere. Moreover, the plasma and magnetic field are evolved fully consistently in an MHD system. Last but not least, the simulation presents a strong active region case, which is a more intriguing one in real observations.

To assess the force-freeness of the magnetic field, we use the following factor:
\begin{equation}
r_{f} = \frac{\lvert \mathbf{j} \times\mathbf{B}\rvert}{\lvert\mathbf{j}\rvert\cdot\lvert\mathbf{B}\rvert},
\label{equ:fff}
\end{equation}
which is the ratio between the current perpendicular to the magnetic field and the full current. Obviously, $r_{f}{=}0$ represents a force-free field, where currents are completely field-aligned, and the larger $r_{f}$ is, the more non-force-free the magnetic field becomes. The angle between $\mathbf{j}$ and $\mathbf{B}$ is given by $\arcsin(r_{f})$. 

In each horizontal layer through the whole vertical domain from $z{=}-9.6$\,Mm to $z{=}113$\,Mm, we calculate the histograms of $\log_{10}\beta$ and $r_{f}$. The former is done for a range between -6 and 6 with an interval of $\Delta\,\log_{10}\beta{=}0.1$, and the latter uses an interval of $0.01$. \figref{fig:ar_forcebeta} displays the probability density functions (PDFs) of $\log_{10}\beta$ and $r_{f}$. In particular, we also plot detailed views of the domain between $z{=}-5$\,Mm and $z{=}5$\,Mm, where $\beta$ and $r_{f}$ change the most significantly. We note that the histogram at each height is normalized to unity.

We first look at the gas-dominating domain below the photosphere.  The PDF of $r_{f}$ concentrates to the right of $r_{f}=0.6$ and quickly vanishes as $r_{f}$ approaches zero. The (most probable) mean values of $r_{f}$ are around 0.9 in this region, indicating a highly non-force-free condition. Panel (c) shows that almost all grid points below $z{=}0$\,Mm fall into the high $\beta$ side, and the mean $\beta$ is over $10^4$ for most of the convective layers. Only in the layer very close to the photosphere can we see the PDF of $\beta$ extending to the low $\beta$ side, which is probably due to the strong magnetic flux concentrations, where the magnetic pressure prevails over the gas pressure. The highly non-force-free state of the magnetic field is expected in the convective layer, where the magnetic field is constantly stressed by turbulent convection and cannot relax. In this passive state of the magnetic field, the magnetic field vector direction is random relative to the direction of the current vector, and there is a far larger solid angle span for the magnetic field vector to be perpendicular to the current direction than aligned with it. Hence, the PDF peaks at $r_f$ close to one.

The observation-relevant and the most interesting region is the domain above the photoshere, where both $r_{f}$ and $\beta$ exhibit the opposite trend compared with the convection zone.  The zoomed view in \figref{fig:ar_forcebeta}(b) shows that the mean $r_{f}$ jumps from 0.9 to around 0.1 across a thin transition layer near $z{=}1$\,Mm. Above this layer, the mean $r_{f}$ remains at similar values until $z{\approx}90$\,Mm. This indicates that many current vectors are at a small but non-vanishing angle with respect to the magnetic field vectors in the corona, and hence the coronal magnetic field is essentially non-force-free. The distribution of $r_{f}$ peaks at 0.09 ($\sim5^{\circ}$) in the lower corona, with a non-negligible wing on the right of the peak. For example, at z=40\,Mm, the probabilities of $r_{f}{\leq}$0.18 ($10.4^{\circ}$), 0.35 ($20.5^{\circ}$), and 0.6 ($36.9^{\circ}$) are 0.45, 0.70, and 0.90, respectively. The PDF of $\beta$ also changes drastically from the high $\beta$ side to the low $\beta$ side. \figref{fig:ar_forcebeta}(d) shows that the mean $\beta$ quickly drops to $10^{-3}$ -- $10^{-2}$ in the first few Mm above the photosphere, and the lowest possible $\beta$ obtained in this snapshot (almost $10^{-5}$) is also found in this region. The mean $\beta$ gradually increases above this height and remains below unity up to $z{\approx}90$\,Mm. As we have noted in \sectref{sec:model}, the mass flux boundary condition strongly dampens vertical motions and leads to a partial trapping of mass and enhancement of gas pressure  near the top boundary (above $z{\approx}100$\,Mm). Consequently, the mean $\beta$ becomes larger than one and the mean $r_{f}$ also demonstrates a clearly non-force-free condition near the top of the domain.

\begin{figure}
\center
\includegraphics{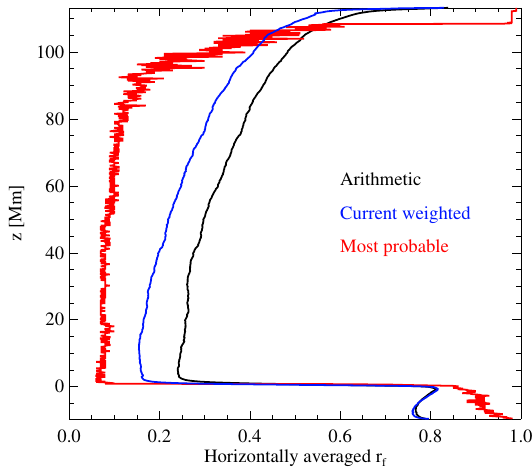}
\caption{Horizontally averaged $r_{f}$ as a function of height. The black line shows the simple arithmetic mean at each height. The blue line plots the current-weighted mean. The red line is the most probable values of $r_{f}$, which is the white line shown in \figref{fig:ar_forcebeta}(a).
\label{fig:ar_jwrforce}} 
\end{figure}

Finally, we compare in \figref{fig:ar_jwrforce} the most probable value of $r_{f}$ as displayed in \figref{fig:ar_forcebeta} and  the current-weighted (CW) mean of $r_{f}$, which is evaluated by 
$$r_{f_{\rm CW}}=\frac{\sum\lvert\mathbf{j}\rvert r_f}{\sum\lvert\mathbf{j}\rvert}$$ 
at each height. The latter is often used to evaluate the force-freeness of (extrapolated) coronal magnetic fields \citep[see ``CW$\sin\theta$" in][]{Wheatland+al:2000,Schrijver+al:2006}. We also plot the simple arithmetic mean of $r_{f}$ as a reference. As we have seen in \figref{fig:ar_forcebeta}, the PDF of $r_{f}$ is highly asymmetric with respect to its peak. Similarly, \figref{fig:ar_jwrforce} shows that the arithmetic and current weighted means clearly deviate from the most probable value. When we compare the simple arithmetic mean and the current-weighted mean, they are very similar below $z{=}0$\,Mm, which suggests that the current is about evenly distributed in all $r_{f}$. However, above $z{=}0$ (in particular the thin transition layer), the arithmetic and current-weighted means become substantially different. The current-weighted mean is much smaller than the arithmetic mean (by a factor of 0.5 -- 0.8). This means that stronger currents are more aligned (i.e., making a smaller angle) with the magnetic field. The current-weighted mean of $r_{f}$ is about 0.15 ($8.6^{\circ}$) in the lower corona below $z{=}30$\,Mm, and gradually increases to around 0.2 ($11^{\circ}$) in the middle part and 0.4 ($23^{\circ}$) near $z{=}100$\,Mm. 

The distribution of plasma $\beta$ in this simulation is consistent with earlier inferences by \citet{Gary:2001,Rodriguez+al:2019} and the MHD coronal simulation by \citet{Bourdin:2017}. Furthermore, we find noticeable deviations from force-freeness. Magnetic field extrapolations that impose force-freeness may also obtain solutions with non-vanishing $r_{f}$ \citep{DeRosa+al:2009}, probably due to numerical inaccuracy. However, the properties of $r_{f}$ in this simulation are a physical result that magnetic forces need to balance the plasma pressure gradient and the viscous stress that dissipates the magnetic energy released through Lorentz force work and heats the coronal plasma.  This suggests the important role of plasma forces for a more accurate modeling of the magnetic field on the real Sun \citep[e.g.,][]{ZhuXiaoshuai+Wiegelmann:2018}.

\section{RESULTS III: Coronal Dynamics Illustrated by Emission Measures}\label{sec:res_em}
Although intriguing dynamics spread over the whole evolution, we select three short time periods to briefly illustrate features and dynamics in the corona. The first period covers the time when the active region magnetic flux first appears in the photosphere and active regions start to form. The second period is in the early stage of active region formation, where newly emerged magnetic flux drives a drastic reorganization of coronal magnetic field on large scales. The third period is in the later half of the evolution, when active regions are well developed.

\subsection{Coronal Dimming above Emerging Active Regions}\label{sec:res_em_dimming}

\begin{figure*}
\center
\includegraphics{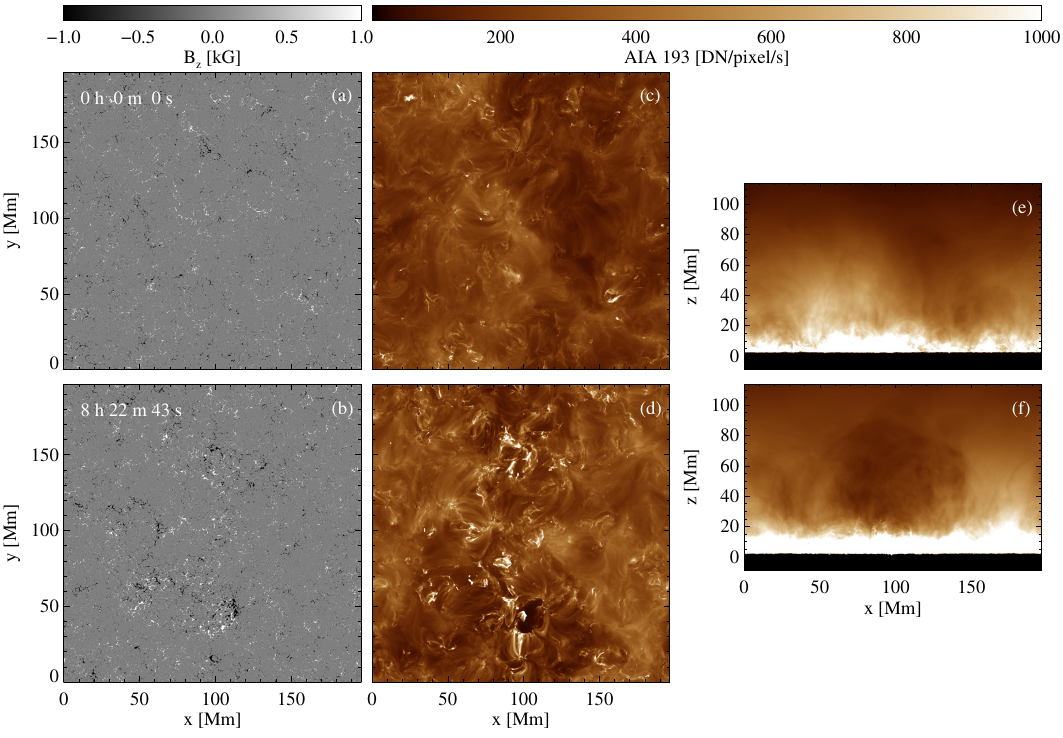}
\caption{Coronal dimming created in the quiet Sun by an emerging active region. (a): $B_{z}$ at the photosphere, before the active region emerges. (b): $B_{z}$ as in Panel (a), but when a flux concentration (at (x,y)=(120,50)) starts to form. (c) \& (d): Synthetic AIA 193 channel images from the top view. The time stamps are the same as in (a) and (b), respectively. The dimming region is visible in the lower half of the field of view. (e) \& (f): Synthetic AIA 193 channel images from a side view along the $y$-axis. The time stamps are the same as in (a) and (b), respectively.
\label{fig:ar_dm_baia}} 
\end{figure*}

We investigate how the first active region that emerged in the quiet Sun changes the corona. \figref{fig:ar_dm_baia} presents a comparison of the photospheric magnetogram and coronal EUV images before (i.e., the quiet Sun) and after the first emerging active region appears in the photosphere. The magnetic field in the quiet Sun is a mixed-polarity field maintained by a small-scale dynamo in the convective layer of the simulation domain. The flux imbalance noticeable on spatial scales of about 20 to 30\,Mm is a manifestation of the convection cells on supergranulation scales. The corona above the quiet Sun, as we have seen earlier in this paper, shows very smooth structures on large scales, with embedded small-scale bright points. 

Panels (b) and (d) in \figref{fig:ar_dm_baia} show the magnetogram and synthetic AIA 193 channel image when the first active region appears. A flux bundle spanning from $x{\approx}50$ to 110\,Mm at $y{\approx}50$\,Mm breaks through the photosphere in the area. The flux concentration at $(x,y){\approx}(120,50)$, which is still rather small in size, already has a field strength well above 1\,kG. The difference in the corona is more evident. The EUV count rate in a small area immediately above the forming sunspots is strongly reduced from a few hundred to almost zero. More interestingly, we can see around the forming sunspots a large area that becomes significantly dimmer than the quiet Sun. This dimming region is over 100\,Mm wide along the $x$-axis. The presence of this dimming region is clearer when comparing Panels (e) and (f). In particular Panel (f) demonstrates that the dimming region is similar to a large bubble that already reaches 80\,Mm above the photosphere at this moment. This implies that even in the very early stage when the sunspots seem to be very small, the emerged magnetic field can already impact a much larger domain.

\begin{figure*}
\center
\includegraphics{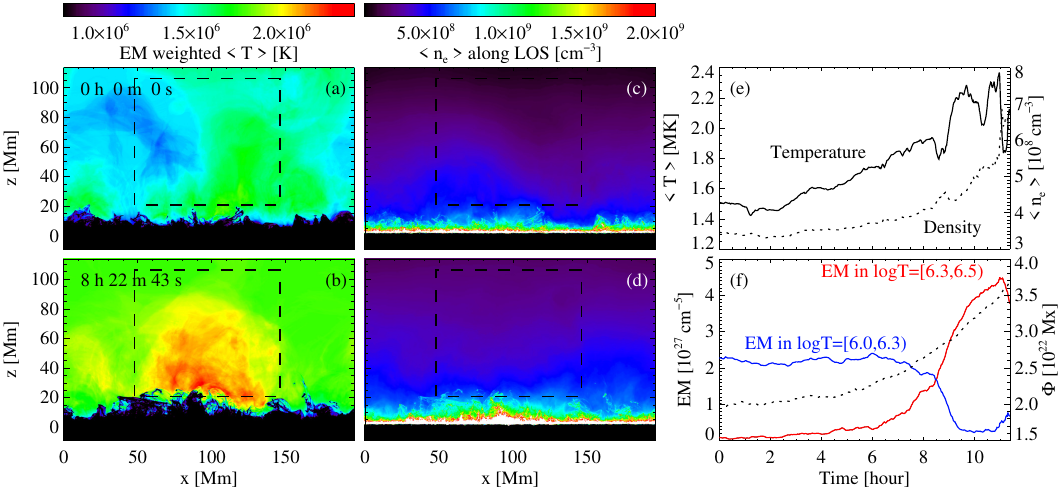}
\caption{Evolution of coronal thermodynamics in the dimming region over an emerging active region. (a) \& (b): Emission measure weighted temperature from a side view along the $y$-axis, as defined in \equref{equ:emwt}. The time stamps are the same as in \figref{fig:ar_dm_baia}. (c) \& (d): Mean density derived from emission measures, as in \equref{equ:emwn}. (e): Change in the mean temperature and density of the corona above the emerging active region (averaged within the boxes marked in Panels (a)--(d)). The solid and dotted lines show the mean temperature and density, respectively. (f): Change in the mean emission measure (EM) in the corona above the emerging active region (averaged within the box marked in Panels (a)--(d)). The blue line shows the mean emission measure in $\log T{=}[6.0,6.3)$, and the red line is for $\log T{=}[6.3,6.5)$. The black dotted line shows the unsigned magnetic flux in the photosphere.
\label{fig:ar_dm_rtem}} 
\end{figure*}

The dimming region associated with an expanding magnetic field shows some similarity to dimming regions associated with CMEs, which are caused by the evacuation of mass \citep[e.g.,][]{ChengXin+al:2012,Dissauer+al:2018}. However, this is {\it not} the case for the dimming region here. We use emission measures to investigate the change in plasma properties during this period.

 We evaluate the emission measure weighted logarithmic temperature by
\begin{equation}\label{equ:emwt}
\langle \log T \rangle = \frac{\sum \limits_{n=0}^{N_{em}-1}EM(\log T_{n})\,\log T_{n}}{\sum \limits_{n=0}^{N_{em}-1}EM(\log T_{n})}.
\end{equation} 
Alternatively, we can equivalently define a mean temperature by $\langle T \rangle{=}10^{\langle \log T \rangle}$. In either form, it represents the temperature averaged along a line of sight and reflects the temperature of plasma that contributes the most emission. 

The sum of emission measures over all temperatures represents the integral of $n_{e}^{2}$ along the line of sight. Therefore, we can estimate the mean density, $\langle n_{e} \rangle$, by
\begin{equation}\label{equ:emwn}
\langle n_{e} \rangle = \sqrt{\sum \limits_{n=0}^{N_{em}-1}EM(\log T_{n})\bigg/L},
\end{equation}
where $L$ denotes the length of the line of sight.

The maps of $\langle T \rangle$ and $\langle n_{e} \rangle$ for a side view along the $y$-axis are presented in \figref{fig:ar_dm_rtem}. Panels (a) and (b) compare $\langle T \rangle$ before and after the active region appears.  Panels (c) and (d) illustrate the change in $\langle n_{e} \rangle$. For the region above the forming active region marked by the boxes in \figref{fig:ar_dm_rtem}, we plot the mean temperature and density as functions of time in \figref{fig:ar_dm_rtem}(e), and Panel (f) plots the temporal evolution of coronal emission measures in two adjacent temperature ranges.

The mean temperature of the quiet Sun corona is around 1.5\,MK and there is a large area with even cooler temperatures. When the active region appears, the most evident feature in the mean temperature is a dome of about 2\,MK. This high temperature dome extends over 100\,Mm in the horizontal direction and reaches 80\,Mm above the photosphere, which exactly matches the dimming region seen in the synthetic EUV images. Meanwhile, the mean temperature in the ambient corona is also slightly increased. This phenomenon is clear evidence that emerging active regions provide a significant excessive energy flux that heats the quiet sun corona to higher temperatures, and this heating is already present and impacts the large-scale corona in the very early stage of active region formation.

 The mean density of the quiet Sun corona increases gradually from $2\times10^{8}$ to $6\times10^{8}$\,cm$^{-3}$ as the newly emerged active region pushes plasma upward from the lower atmosphere. The mean density  below $z{=}20$\,Mm (i.e., below the box marked in \figref{fig:ar_dm_rtem}) is increased significantly. We can also discern a very slight decrease in the density in some areas at $z{\approx}60$\,Mm, which implies that the expanding magnetic field is reorganizing pre-existing plasma in the corona, and perhaps it is driving the plasma sideways. 

The changes in the mean temperature and density are consistent with the expectation that the forming active region gives rise to more intensive heating and pushes masses into the corona. The most important information is given by the emission measure. In the first seven hours, the coronal emission measure is mostly in the temperature range of $\log T{=}[6.0,6.3)$. However, the emission measure in a higher temperature range keeps growing with the growth of magnetic flux in the photosphere. During a period of one hour at $t{\approx}8.5$\,h, the hotter emission drastically increases, and the cooler emission significantly drops.

To conclude, the coronal dimming region found in the early stage of active region formation is a manifestation of intense heating that evolves the quiet Sun corona into an active region corona. So-called dark moat, a dimmed area around {\it developed} isolated active regions, has been found for decades \citep[see ][and references therein]{Singh+al:2021}. Recent observations revealed that {\it emerging} active regions give rise to emerging dimming regions \citep{ZhangJun+al:2012,Payne+SunXudong:2021} or a large-scale outwards-propagating dark ribbon \citep{Zhang+al:2015}. The response of the quiet Sun to the emerging active region in this simulation is consistent with the phenomenon observed by \citet{Payne+SunXudong:2021}, particularly the evolution of the emission measure in the dimming region, and confirms that the emerging dimming regions are caused by enhanced heating in the corona.

\subsection{EUV Emissions above Emerging Active Regions}\label{sec:res_em_early}

\begin{figure*}
\center
\includegraphics{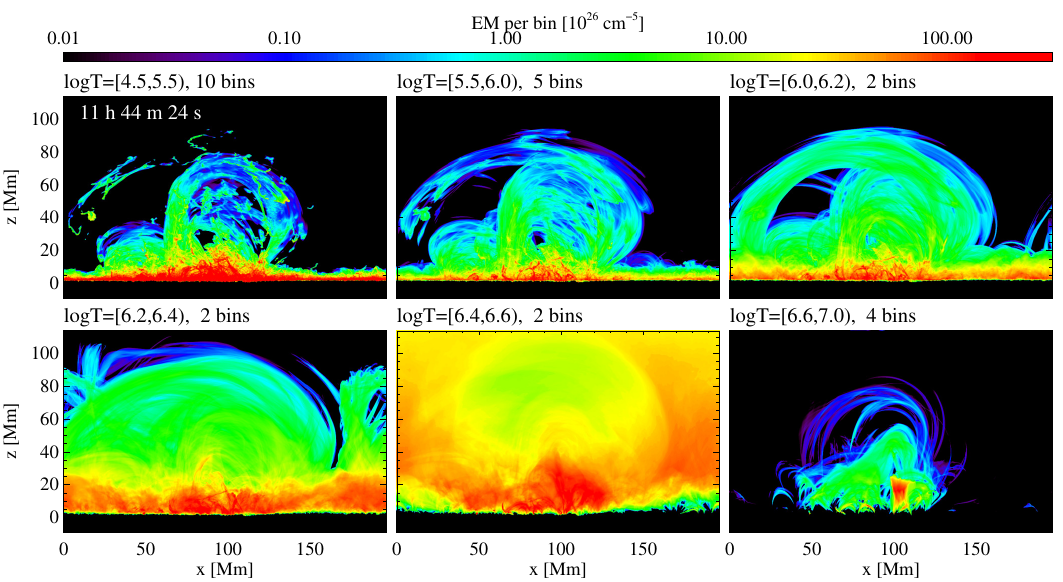}
\caption{Coronal emission measure in temperatures from $\log T{=}4.5$ to $\log T{=}7.0$ when cool plasma in the lower atmosphere is ejected into the corona after an eruption occurs in the early stage of the simulation. Each panel displays the total emission measure in a selected temperature range divided by the number of bins it contains (with a bin interval of $\Delta\log T{=}0.1$). The corresponding temperature range and number of bins are noted above each panel. This figure is a snapshot of an animation that covers about 2.5 hours of evolution (available in the electronic version of this paper). The animation presents how the ejected plasma falls back to the lower atmosphere, as well as the formation of many long coronal loops, which is described in more detail in \sectref{sec:res_em_early}. 
\label{fig:ar_zem_early}} 
\end{figure*}

In \figref{fig:ar_zem_early} and the associated animation, we present a side view (along the $y$-axis) of the coronal emission measure in a time period of about 2.5 hours starting at about $t{=}11\,$h. This period is in the early stage of the full active region evolution, when fast flux emergence is ongoing and active regions start to form. The emission measure during this time period highlights a large-scale eruption of cool material from the low atmosphere and the consequences of this eruption, such as the formation of large-scale coronal loops and the enhancement of coronal temperature.

The emission measure calculated at runtime has an interval of $\Delta {\rm log}T=0.1$ and covers temperatures ranging from $\log T{=}4.5$ to the highest temperature in the domain. For presentation in this section, we rebin the emission measure in neighboring bins and reform the emission measure array to fewer bins. \figref{fig:ar_zem_early} is a snapshot of the animation at $t{=}$11\,h\,44\,m and presents the emission measure that has been rebinned into six bins. This snapshot captures a large-scale eruption, which is the most interesting highlight in this time period.

\subsubsection{large-scale Eruption}
The animation shows that the eruption starts at $t{=}$10\,h\,59\,m, when the emission in the hottest bin suddenly increases and a hot loop appears at $(x,z){\approx}(80,10)$. Immediately after that time, a large amount of cool plasma ($\log T{<}6.0$) is ejected from the low atmosphere and reaches over 80\,Mm in height in about 6 minutes. Large parcels of ejected plasma quickly break into small fragments that exhibit various evolution. The falling plasma behaves like coronal rains \citep[e.g.,][]{Schrijver:2001,Antolin+al:2012,Antolin+al:2015,DePontieu:2021.IRIS} and traces the magnetic field lines. However, we also see many fragments that seem to ``float'' in the corona and can dynamically move up and down. Their movement illustrates the dynamic evolution of the magnetic field over emerging active regions, which was not considered by other numerical models of coronal rains \citep[e.g.,][]{Mueller+al:2005,Xia+al:2017,Kohutova+al:2020}. Moreover, the falling plasma may undergo a cooling process as in a typical coronal rain scenario or a heating process that converts the cool plasma into a mass source to hotter coronal loops.

Magnetic field lines with a dip were suggested to be the favorable geometry for filament formation \citep{Mackay+al:2010}, which has been tested by numerical experiments \citep[e.g., ][]{XiaChun+Keppens:2016}. The magnetic field in the corona of our simulation does not form a dip. Otherwise, we could very likely see a filament/prominence formed by these falling cool plasma elements.

\subsubsection{Fast Expanding Coronal Loops}
In a period of about 20 minutes after $t{=}11$\,h\,20\,m, the falling plasma gradually fills a large magnetic arcade rooted at $x{\approx}0$ and 130\,Mm, and this forms a large coronal loop mostly visible around $\log T{=}6.0$. If we compare the panels of $\log T{=}[5.5,6.0)$ and $\log T{=}[6.0,6.2)$, patches that disappear in the former will appear in the latter. After $t{=}12$\,h, the large loops almost disappear in the bin of $\log T{=}[5.5,6.0)$ but remain visible in the bin of $\log T{=}[6.2,6.4)$. Meanwhile, another group of loops emerges in the area between $x{\approx}70$ and 130\,Mm. They first become visible in the lower temperature bins ($\log T{<}6.0$) at about $t{=}11$\,h\,30\,m and outline magnetic field lines that are expanding upward. At $t{=}12$\,h, these loops completely fade away in the bin of $\log T{=}[4.5,5.5)$ and become more evident in the bin of $\log T{=}[6.0,6.2)$. About 35 minutes later, they can only be seen in bins of $\log T{>}6.0$ and have expanded to a height of about 100\,Mm. Finally, these loops disappear in the bin of $\log T{=}[6.0,6.2)$ and are only visible in the bin of $\log T{=}[6.2,6.4)$. 

The shift of similar structures from bins of lower temperatures to those of higher temperatures, as illustrated by the two examples here, provides clear evidence of plasma heating during fast flux emergence. The energy input can efficiently heat the cool plasma carried in emerging magnetic field lines to temperatures well above one MK and maintain the density at loop apexes for a long time. Another interesting behavior of loops formed in the fast emergence stage is that they are constantly changing. As we see in the animation, a loop or thread that is bright in a certain bin may fade and disappear in a few minutes, and a similar feature is very likely to be found in a bin of higher temperatures. Furthermore, if we observe a structure that seems to remain at one place, this does not necessarily imply a static state. However, it is probably a combined effect of old threads disappearing (moving to higher temperatures) and new threads forming (heated from lower temperatures).

\subsection{EUV Emissions above Developed Active Regions}\label{sec:res_em_later}

\begin{figure*}
\center
\includegraphics{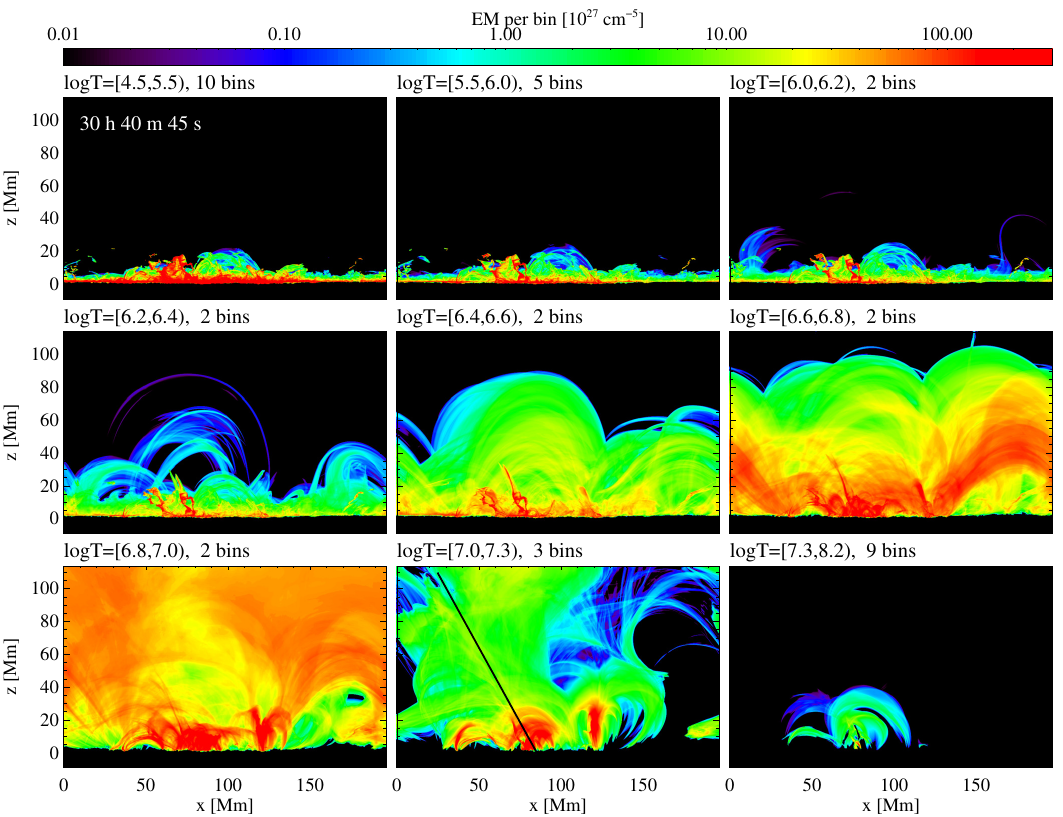}
\caption{Coronal emission measures at temperatures from $\log T{=}4.5$ to $\log T{\geqslant}8.0$\ when a high speed jet occurs in a late stage of the simulation. The quantity shown in each panel is the total emission measure in the temperature range divided by the number of bins (with a bin interval of $\Delta\log T{=}0.1$), which are noted above each panel. The dotted line in the panel of $\log T{=}[7.0,7.3)$ indicates the approximate direction of the propagation of the jet. This figure is a snapshot of an animation that covers about one hour of evolution (available in the electronic version of this paper). In addition to the onset of the jet, the animation also highlights the evolution of a set of cooling coronal loops gradually appearing in the $\log T{=}[6.2,6.4)$ panel and transient and stable hot compact loops $\log T{\geqslant}6.8$ in the active region core, as describe in \sectref{sec:res_em_later}.
\label{fig:ar_zem_later}} 
\end{figure*}

\begin{figure*}
\center
\includegraphics{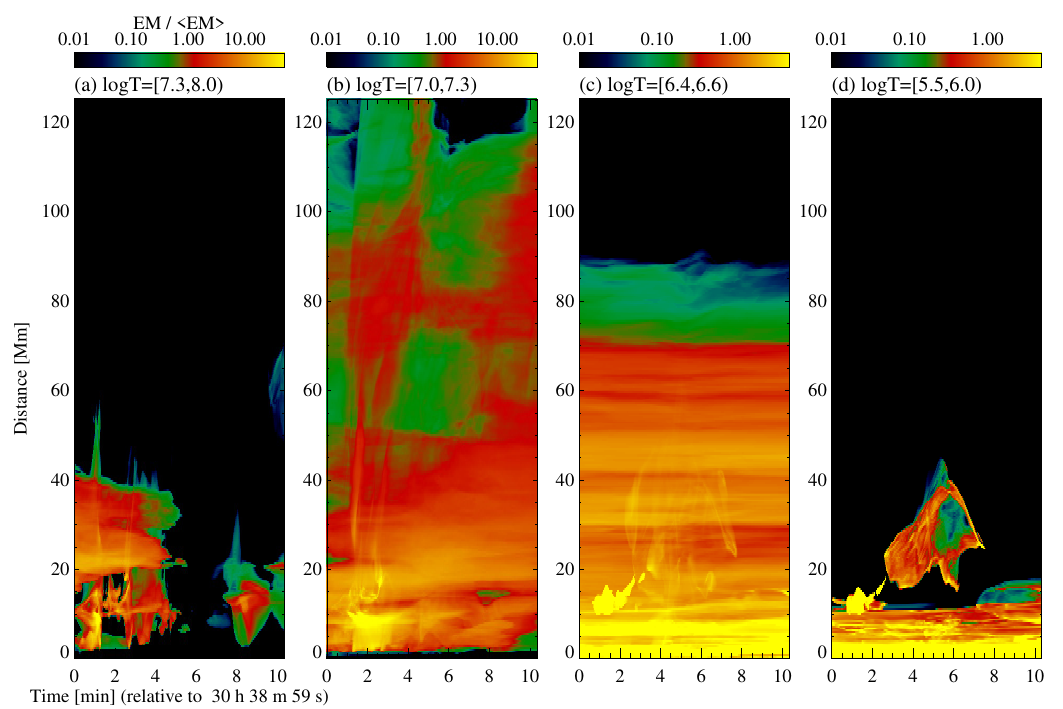}
\caption{ Time-distance diagrams of emission measures in four temperature bins showing the onset, propagation, and decay of the jet seen in \figref{fig:ar_zem_later} and associated animation. The diagram is constructed for the duration of the jet (about 10 minutes) and along the dotted line in \figref{fig:ar_zem_later}. The EM in each panel is shown relative to the mean value.
\label{fig:ar_zem_td}} 
\end{figure*}

In \figref{fig:ar_zem_later} and the associated animation, we present a side view (along the $y$-axis) of the coronal emission measure in a late stage of the full evolution, when flux emergence in the photosphere slows down and active regions are well developed. The time period presented here starts at about $t{=}30.5$\,h and lasts for about one hour. The emission measure features a high-speed jet propagating through the vertical domain and dynamics of typical coronal structures above developed active regions, such as small-scale surges of cool plasma, warm coronal loops arching over active regions, and hot compact loops in active region cores. 

In this time period,  coronal plasma is commonly seen in the temperature range between a few MK to 20\,MK, whereas a small amount of hot plasma may transiently reach temperatures of several tens MK. The hottest plasma is found in regions of flares that occur intermittently. Therefore, to accommodate the much higher temperature maximum (compared with \figref{fig:ar_zem_early}), we reform the original output of the emission measure to nine temperature bins and present a snapshot of the rebinned emission measure at $t{=}$30\,h\,40\,m in \figref{fig:ar_zem_later}. This figure is also a frame of the animation showing the temporal evolution during this time period.

\subsubsection{Fast Propagating Hot Jet}
Jets are ubiquitously observed on the Sun from the chromosphere to the corona \citep{Raouafi+al:2016,ShenYuandeng:2021}. An intriguing event seen during this period is a fast propagating jet coming out from the core of the active region. The jet is first seen in the hottest bin of $\log T{\geqslant}7.3$. It can be identified in frames around $t{=}30$\,h\,39\,m\,40\,s as a thin thread along a line pointing from $(x,z){\approx}(80,0)$ to $(x,z){\approx}(60,60)$. The jet quickly disappears in the hottest bin in less than 20 seconds and subsequently appears in the bin of lower temperatures.

A snapshot of the animation at $t{=}30$\,h\,40\,m\,45\,s is displayed in \figref{fig:ar_zem_later}. At this moment, we can see in the bin of $\log T{=}[7.0,7.3)$ a trace that extends from its onset location to the upper left corner of the field of view,  which is roughly marked by the black line in \figref{fig:ar_zem_later}. In the animation, the trace shows up almost instantaneously over a large distance, yielding an apparent propagation speed well over 1000 km/s. Thus, it would be very interesting to clarify by a further analysis (not performed in this study) whether this is an actual flow or a heating front. Meanwhile, the lower part of the jet can be observed in bins of intermediate temperatures (middle row in \figref{fig:ar_zem_later}) with a similar shape and propagation direction as it first appeared in bins of much higher temperatures. The propagation speed of the structure seen in these temperatures is about 500\,km/s, which is a fast event compared with observed EUV/X-ray jets \citep{Shimojo+al:1996,Savcheva+al:2007}.

The animation shows that in a few minutes after the time instance shown in \figref{fig:ar_zem_later}, the jet gradually cools down and becomes visible at lower temperatures (bins of $\log T{<}6.0$). Meanwhile, the jet also expands in the transverse direction, thus, the structure seen in the cool bins is significantly wider than that seen in the hot bins. Eventually, after about $t{=}30$\,h\,45\,m, the ejected plasma gradually falls back down.

 To better illustrate the fast transient discussed above, we construct time-distance diagrams of the emission measure over the duration of the jet and along the black line marked in \figref{fig:ar_zem_later}. \figref{fig:ar_zem_td} shows the T-D diagrams in four temperature bins. The jet-related features are first seen in the high temperature bins, and in particular, the jet induces a spatially extended impact in the bin of $\log T{=}[7.0,7.3)$. The large slope of the trace also indicates the high propagation speed (either apparent or actual) of the impact. In the bins of lower temperatures, we can see the trajectory of plasma rising and falling.

Jets triggered by flux emergence have been extensively studied by numerical simulations in recent decades\citep[e.g.,][and references therein]{Yokoyama+Shibata:1995,Archontis+al:2005,Moreno-Insertis+Galsgaard:2013,Fang+al:2014,Iijima+Yokoyama:2017}. The jet seen in our simulation occurs in an unprecedentedly complex environment compared with previous simulations. It is interesting to study how the configuration of the ambient magnetic field affects the generation and propagation of the jet, as suggested by \citet{Pariat+al:2015}.

\subsubsection{Warm Coronal Loops}
Loops are the most common feature observed in the corona \citep{Reale:2014}. The middle row of \figref{fig:ar_zem_later} shows prominent coronal loops that are observed in intermediate temperatures (warm loops) as arches reaching tens of Mm over the solar surface. At the beginning of the animation, we can see several loops with apexes at $z{\approx}40$ to 60\,Mm in the bin of $\log T{=}[6.2,6.4)$. At the same time, it is more difficult to isolate a single loop from the bin of $\log T{=}[6.4,6.6)$, but there are many thin threads above the rather diffusive background. In the bin of $\log T{=}[6.6,6.8)$, we can identify a wide loop-like arch that roots at $x{\approx}45$ and 125\,Mm and reaches about $z{\approx}$70\,Mm at the apex. 

In the following 30 minutes, the loop-like arch in the hotter bin in the middle row slightly decays, and several bright loops gradually appear in the cooler bin. At $t{=}31$\,h, these loops can be seen in the cooler bin as nice arches with apexes at $z{\approx}$60 to 80\,Mm and feet separations of about 80\,Mm. These geometric properties are consistent with the structure that was identified earlier in the hotter bin. All these signatures provide clear evidence that the new loops seen in the cooler bin result from pre-existing structures cooling down from higher temperatures. In the next 30 minutes after $t{=}31$\,h, the loops vary constantly as existing bright threads may disappear and new threads also form. However, the overall structures in the bins in the middle row of \figref{fig:ar_zem_later} do not change significantly. 

The cooling behavior is very often observed in real coronal loops \citep[e.g,][]{Schrijver:2001,Winebarger+al:2003,Winebarger+Warren:2005,Viall+Klimchuk:2012,LiLeping+al:2015,Froment+al:2015}. This has been suggested as evidence of impulsive heating in coronal loops and an explanation for warm loops whose observational properties are inconsistent with predictions of hydrostatic equilibrium \citep{Aschwanden+al:2001,Warren+al:2002,Warren+al:2003,Klimchuk+al:2010}. In our simulation, non-equilibrium is the most common state of coronal loops over actively evolving active regions.

\subsubsection{Hot Loops in Active Region Cores}

As the last part of this section, we examine the emission from the hottest plasma in the corona, which is presented in the bottom row of \figref{fig:ar_zem_later}. We first see a significant amount of emission by hot plasma in the high corona, as shown in the bin of $\log T{=}[6.8,7.0)$. This is actually caused by the top and (periodic) side boundaries that do not allow for energy flux effectively leaving the domain. Meanwhile, a very high energy input in the corona is provided by the strong active regions that already contain $10^{23}$\,Mx of unsigned flux at this moment. The active regions in this simulation are comparable to the largest active regions seen in the real Sun in terms of magnetic flux, and real active regions would not be confined in a limited volume as in this simulation. Therefore, the superhot emission in the high corona near the top of the simulation domain may not fully represent the real corona.

In the lower corona, short coronal loops with apexes generally below 20\,Mm give rise to significant emission measures at temperatures of about $10^{7}$\,K. These hot loops can be stable or transient structures. For instance, the stable structure appearing as a vertical ``plume" at $x{=}120$\,Mm in the bin of ${\rm log}T{=}[7.0,7.3)$ is a loop that extends mostly along the $y$ direction (i.e., the line-of-sight for this figure) and connects two strong sunspots at $y{\approx}60$ and 90\,Mm. This loop is formed between the two sunspots and remains hotter than 10\,MK until the end of the evolution. The animation of \figref{fig:ar_zem_later} shows many transient hot loops with lifetimes ranging from a few seconds to tens of minutes. 

These hot loops in the active region cores resemble the hot plasma revealed by X-ray observations of solar active regions \citep{Schmelz+al:2009,Reale+al:2009,Warren+al:2012}. The active region cores in the simulation appear to be highly structured and are surrounded by more diffusive features, which is similar to the observation reported by \citet{Reale+al:2007}. The presence of short-lived and steady features is also found in observation \citet{Warren+al:2010}. 

Whether impulsive or steady heating dominates in active region cores is not fully clear in observations \citep[see, e.g.,][and references therein]{Tripathi+al:2011,Warren+al:2010,Winebarger+al:2011}. However, it is very likely that both mechanisms work in our simulation. \citet{Chitta+al:2018} proposed that the cancellation of magnetic fields in opposite polarities can provide sufficient energy for hot loops. This could be a viable mechanism for the transient hot loops in this simulation. In particular, the ongoing flux emergence continuously brings small patches of magnetic field that can trigger cancellation with a pre-existing magnetic field. However, the long-lived and steady hot loops that last for over 20\,hours are more likely to be maintained by (statistically) steady heating. This leaves interesting open questions for more in-depth analysis of the energy release in the active region cores.

\section{SUMMARY}\label{sec:sum}
We presented the first results of a realistic radiative MHD simulation of the Sun's uppermost convection zone and atmosphere. The control equations of this simulation consider radiative transfer in the lower solar atmosphere and take into account optically thin radiation and field-aligned thermal conduction in the corona. This allows for realistic modeling of thermodynamic properties of coronal plasma, hence a direct and quantitative comparison between synthetic observables and actual observations. The large vertical domain spanning from the uppermost convection zone to over 100\,Mm in the corona allows this simulation to properly account for the coupling between difference layers of the solar atmosphere though magnetic field and mass and energy flows in a single simulation domain. We studied the self-consistently maintained quiet Sun corona and the formation of strong solar active regions through magnetic flux emergence. The results presented in this paper are summarized as follows.

\subsection{General Properties of the Quiet Sun and Active Regions}
The quiet Sun run reaches a dynamic equilibrium, where the vertical profiles of temperature and density are consistent with observations. In particular, the mean coronal temperature, which remains stable at about 1.5\,MK, is maintained by dissipating the Poynting flux that is naturally generated by the magneto-convection at the solar surface. In addition to the stable background quiet Sun corona, we also demonstrated various small-scale activity in the quiet Sun. The most evident ones are the numerous EUV bright points seen in the synthetic observations and the propagating disturbances, as revealed by the running difference ratio of synthetic AIA images.

The active region run considers the emergence of magnetic flux from the uppermost convection zone to the corona and the formation and evolution of solar active regions in a time period of about 50 hours. The setup of the active-region scale magnetic flux is a key ingredient of the active region simulation. Unlike the idealized flux tubes/sheets that were extensively used in previous studies, our simulation adopts the flux bundles generated in a global-scale solar convective dynamo \citep{Fan+Fang:2014}. The significantly improved realism in the magnetic field leads to more realistic  sunspot groups that can reproduce the asymmetries of observed sunspots, as discussed in Paper\,I. Moreover, these flux bundles also give rise to complex active regions that turn out to be very flare productive. The eruptions that occur in the simulation will be presented in a forthcoming separate paper.

The results on the evolution of the sunspots seen in the photosphere are consistent with those presented in Paper\,I. These include the emergence of magnetic flux bundles to the photosphere, the formation of two major sunspot pairs (marked as C1 \& C2 and S1 \& S2 in Paper\,I), and the asymmetry that the leading spots are stronger and more coherent. The simulations in Paper\,I employed the same setup but with the top boundary placed at about 600\,km above the photosphere. In comparison, the top boundary of the present simulation is sufficiently farther away from the photosphere. Therefore, we can confirm that the results in Paper\,I were not (top) boundary effects. In both studies, the behavior of the sunspots in the photosphere is dominated by the imposed emerging flux bundles and convective flows from the dynamo simulation.

\subsection{Coronal magnetic field and its energy partition}
We focused more on the evolution of the coronal plasma and magnetic field above the active regions in the present study. The total unsigned photospheric magnetic flux grows to more than $10^{23}$\,Mx, which makes the simulated active regions comparable to the largest active regions observed on the real Sun. The total magnetic energy in the volume of about $197^{2}\times113$\,Mm$^{3}$ increases from $8.5\times10^{30}$\,erg for the quiet Sun to $4.7\times10^{33}$\,erg at the end of the evolution. About 18\% of the total magnetic energy is free energy. This ratio remains relatively constant during the evolution. We investigated the distribution of magnetic energy with height for three evolution stages. The free energy decays slower than the total magnetic energy. As a result, the magnetic field in the upper half of the domain (i.e., $z{>}60$\,Mm) is highly non-potential, in the sense that the free energy contributes 80\% to almost 100\% of the total magnetic energy. 

This active region model gives a {\it realistic} presentation of the force-freeness and plasma $\beta$ in a single domain covering two extreme regimes. The magnetic field and plasma in this simulation evolve consistently from the beginning, and the magnetic force, plasma force, and viscous stress find their balance spontaneously. The implications are twofold. On the one hand, the Lorentz force cannot vanish in the real corona. Even in a purely quiescent area, it is necessary to have a residual Lorentz force to balance the pressure gradient and to release energy as part of the coronal heating process, which implies a balance with acceleration and viscous forces. As shown in our simulation, the current vector is rarely parallel to the magnetic field vector, and the distribution of the angle between the two vectors has an enhanced wing on the right side (larger than) of the most probable value. On the other hand, for most of the corona volume, the angle remains relatively small, which suggests that the force-free field seems to be an acceptable assumption. 

The distribution of the plasma $\beta$ follows the same scenario. Beneath the photosphere, $\beta$ is in the range between $10^2$ and $10^6$, whereas $\beta$ rapidly drops to below unity above the photosphere. The lowest (mean) value of about $10^{-3}$ is seen in the transition region, and the local minimum can be approximately $10^{-5}$. A detailed view of the vertical domain from 5\,Mm below to 5\,Mm above the photosphere shows that the transition from the highly non-force-free and high $\beta$ regime to the (close to) force-free and low $\beta$ regime occurs in a height range of about just one Mm.

\subsection{Dynamics of Coronal Plasma}
We highlighted three time periods in different evolution stages of active regions and demonstrated coronal features that are comparable to structures and dynamics in observations. 

When the first active region starts to appear in the photosphere, we found in the synthetic AIA images a bubble-like dimming region that slowly expands from the active region to almost the whole corona. We investigated the emission measures and thermodynamic properties in the corona above the forming active region. Both the mean temperature and density in the corona increase, and this trend coincides with the growth of the photospheric magnetic flux. Consequently, the emission in the temperature range of the original quiet Sun corona drops by an order of magnitude, which leads to the dimming region. As a comparison, the synthetic images of a filter that is sensitive to higher temperatures would demonstrate a brightening above the forming active region.

We demonstrated two eruption events that occur in the early and late stages of active region evolution. In the early stage when the active regions are forming through fast flux emergence, a large-scale eruption ejects a significant amount of plasma from the lower atmosphere to over 100\,Mm in the corona. This eruption provides an instance of mass circulation on large scales. Ejected plasma with a temperature of about $10^5$\,K may fall back to the solar surface along magnetic field lines or be heated and become a mass source for hotter coronal loops. Some clumps could ``float'' up and down in the corona, as they are supported by the magnetic field. 

A fast propagating jet occurs in the late stage when the active regions have developed strong and complex magnetic fields. The speed of the bulk of the jet is over 500\,km/s, and the {\it apparent} speed of the leading front of the jet exceeds 1000\,km/s. The plasma in the source region of the jet is heated to extremely high temperature (${>}20$\,MK), which is strong evidence of magnetic reconnection.

We highlighted different types of coronal loops through the evolution. In the early stage, dynamic and transient large-scale coronal loops quickly form after an eruption. At the same time, the entire corona undergoes a rapid increase in temperature and evolves from the quiet Sun to a highly structured active region corona. 

In the late stage when active regions are more developed, a group of long and warm loops (1--3\,MK) forms through the cooling of loops that were heated to much higher temperatures (${>}$5\,MK). Individual bright thread within the group of loops appears and disappears in a very short time. However, if we examine the group of loops as a whole structure, it is sustained for a lifetime that is much longer than the cooling time scale. We found both stable and transient hot loops at temperatures of about 10\,MK  in the active region cores. These loops appear to be independent structures and imply the presence of impulsive and statistically stable heating at the same time.
 
\section{Conclusion}\label{sec:con}
In this paper, we present an introduction to the method and the first results of a comprehensive model of quiet Sun and solar active regions. This model provides in a single simulation an unprecedented combination of a computational domain that is wide enough to host large active regions and spans from the uppermost convection zone to the corona, a billion grid point mesh, a long temporal evolution including about 11.5 hours of quiet Sun and 48 hours of active region, strong and complex active region scale magnetic flux, and sophisticated physics for radiative and conduction energy transport in the solar atmosphere, each of which is truly a challenge for numerical simulations.

The scope of the current paper is to present an overview of the general properties of the simulated quiet Sun and active regions, with highlighting a few results that could lead to a wide range of subsequent studies. In addition to the very wide range of topics covered in this paper, more than 170 flares occur in 48 hours of the active region stage. The largest flare reaches M class and is accompanied by a spectacular coronal mass ejection. An overview of the results related to solar eruptions will be presented in a forthcoming paper.

The simulation generated a total of about 400\,TB data, including full 3D snapshots of the whole computational domain, 2D diagnostic slices (plane cuts, $\tau$ surfaces, and emission measures as presented in this paper) at high temporal cadence, and the presented results are from the analysis of less than 10\% of them. The data will be open to the community. However, due to limited manpower and IT resources, these data cannot be made available as a whole package. For the moment, interest of using the data can be addressed to authors of this paper.

\begin{acknowledgements}
We thank the anonymous referee for many helpful suggestions. This material is based upon work supported by the National Center for Atmospheric Research, which is a major facility sponsored by the National Science Foundation under Cooperative Agreement No. 1852977. We would like to acknowledge high-performance computing support from Cheyenne (doi:10.5065/D6RX99HX) provided by NCAR’s Computational and Information Systems Laboratory, sponsored by the National Science Foundation. F.C. is supported by the National Key R\&D Program of China under grant 2021YFA1600504, the Fundamental Research Funds for the Central Universities under grant 0201-14380041, and the Program for Innovative Talents and Entrepreneurs in Jiangsu. F.C. had been supported by the Advanced Study Program postdoctoral fellowship at NCAR and by the George Ellery Hale postdoctoral fellowship at the University of Colorado Boulder since this project was initiated in 2016.
\end{acknowledgements}

\bibliography{reference}
\end{document}